\documentclass[]{article}
\usepackage{fullpage}
\usepackage{graphicx}
\usepackage[colorlinks=true,citecolor=blue,linkcolor=black,urlcolor=blue]{hyperref}
\usepackage{tikz,color,float,epsf,caption,subcaption}
\usepackage{multirow}
\usepackage[style=authoryear, minbibnames=3, dashed=false, backend=bibtex, doi=false, isbn=false, natbib=true,firstinits=true]{biblatex}
\DeclareFieldFormat*{title}{#1}
\renewbibmacro*{volume+number+eid}{%
	\printfield{volume}%
	\setunit*{\addnbspace}
	\printfield{number}%
	\setunit{\addcomma\space}%
	\printfield{eid}}
\DeclareFieldFormat[article]{number}{\mkbibparens{#1}}

\makeatletter
\renewcommand{\fnum@figure}{Fig. \thefigure}
\makeatother

\title{Network science approach for identifying disruptive elements of an airline}
\author{Vinod Kumar Chauhan$^{1,2}$\footnote{This work was done at University of Cambridge UK.}, Anna Ledwoch$^{1,3}$, Alexandra Brintrup$^1$, \\ Manuel Herrera$^1$, Vaggelis Giannikas$^3$, Goran Stojkovic$^4$, Duncan Mcfarlane$^1$\\ $^1$Institute for Manufacturing, University of Cambridge\\ $^2$ Department of Engineering Science, University of Oxford\\
	$^3$ School of Management, University of Bath\\ 
	$^4$The Boeing Company}

\begin{document}
	
	\maketitle
	
	\begin{abstract}
		Currently, flight delays are common  and they propagate from an originating flight to connecting flights, leading to large disruptions in the overall schedule. These disruptions cause massive economic losses, affect airlines’ reputations, waste passengers’ time and money, and directly impact the environment.
		This study adopts a network science approach for solving the delay propagation problem by modeling and analyzing  the’ flight schedules and historical operational data of an airline. We aim to determine the most disruptive airports, flights, flight-connections,  and connection types in an airline network. Disruptive elements are influential or critical entities in an airline network. They are the elements that can either cause (airline schedules) or have caused (historical data) the largest disturbances in the network. An airline can improve its operations by avoiding delays caused by the most disruptive elements.
		The proposed network science approach for disruptive element’ analysis was validated using a case study of an operating airline. The analysis indicates that potential disruptive elements in a schedule of an airline are also actual disruptive elements in the historical data and they should be considered to improve operations. The airline network exhibits small-world effects and delays can propagate to any part of the network with a minimum of four delayed flights. Finally, we observed that passenger connections between flights are the most disruptive connection type.
		Therefore, the proposed methodology provides a tool for airlines to build robust flight schedules that reduce delays and propagation.
		
		\textbf{Keywords :} air transport, flight delays, airline disruptions, delay propagation, network science.
	\end{abstract}
	
	\section{Introduction}
	\label{sec_intro}
	Nowadays, flight delays are increasingly common. In 2017, 20\% flights in Europe arrived late by at least 15 minutes or more (\cite{Walker2017}). These delays can be classified into three main categories: airline, airport, and weather issues. Airline issues arise because of problems at the airline’s end (e.g., delay in passenger boarding and disembarkation, aircraft repairs, and sudden unavailability of crew members). Airport issues arise because of problems at the airport authority’s end (e.g., unavailability of slots for flights to take off or land, longer security checks, and airport closures). The third category includes severe weather conditions that cannot’ be controlled , such as storms and snowfalls, which can disturb airport operations (\cite{brueckner2022airline}).
	
	Airline operations represent a complex distributed transportation system with several interacting and interdependent entities such as passengers, crews, airlines, and airports. Therefore, when one thing goes wrong, it can affect the entire system; thus, delays can propagate to other flights when one flight is delayed. This is called delay propagation, secondary delay, or reactionary delay. In extreme cases, delays can affect hundreds and thousands of flights; for example, \cite{Ledwoch2022} discussed a case in which 2565 flights were affected, and delays were propagated for more than 12 days owing to severe weather conditions. Delay propagation is a major issue for airlines, and it accounted for approximately 30\% to 60\% of delays in European airports in 2017 (\cite{Walker2017}) and approximately 34\% in the U.S. in 2007 (\cite{AhmadBeygi2008}).
	
	``Flight disruption” is a technical term that refers to flight delays and is defined as a “situation where a scheduled flight is cancelled or delayed for 2 hours or more within 48 hours of the original scheduled departure time”\footnote{Airports Council International (ACI), Sixth Worldwide Air Transport Conference (ATCONF), Montreal, 18 to 22 March 2013.}. However, this definition does not represent practical situations adequately. This is because even smaller delays (e.g., a few minutes) can propagate and affect many flights. Moreover, airlines consider a flight delayed only if it is delayed by at least 15 minutes, which is the industry standard for studying the on-time performance of flights. However, even this standard is not practical because  a small delay (say 10 minutes) can cause other flights to be delayed. To understand the root causes and impact of flight delays, it is important to consider even small delays (\cite{Ledwoch2022}).
	
	Flight disruptions have a significant effect on stakeholders and they can be classified into three main categories: airlines and governments, passengers, and the environment. Disruptions cause huge economic losses to the airlines and to the economy of countries. Additionally, they affect airlines’ reputations and passenger retainability. According to Amadaus~(2016), disruptions cost approximately $\$$60~billion per annum  to airlines and their clients globally (\cite{Gershkoff2016}). Flight delays also cause loss of time and money to the passengers and can cause them to miss important events. They also cause environmental pollution by extra emission of harmful gases (i.e., carbon-dioxide ($CO_2$)) owing to fuel combustion. In light of problems of climate change and global warming, this i s a major concern. Thus, considering the significant effect of disruptions on airlines, the environment, and passengers, there is a need to understand the delays and their propagation for airlines, to develop robust schedules to reduce delays and their propagations. \textit{Thus, this study aims to solve delay propagation problem using network science techniques to identify the most disruptive elements of an airline flight network, and enable airlines to build robust flight schedules.} This analysis can help airlines answer the question of which disruptive elements they should focus on improving to enhance their operations.
	
	Currently, networks are ubiquitous, and problems in technology, biology, and social sciences can be formulated as networks (\cite{Newman2018}). This has led to the emergence of a new field, network science,’ (\cite{Brandes2013}). The concepts and techniques of network science can be applied to problems in different domains to better understand them and explain certain phenomena, such as the prevention of infectious diseases, understanding the root causes and effect of flight delay propagation, and studying the extinction of species (\cite{Newman2018}).
	
	Airline networks have also been studied extensively using network science techniques, and data have been modeled as single airline (\cite{Ledwoch2022,Reggiani2010}), multiple airlines (\cite{Baruah2019,Yang2019}), airport (\cite{cumelles2021cascading,Jiang2017,Lordan2019,Wandelt2019}), and flight delay (\cite{Ledwoch2022}) networks. Multilayer networks are recent developments in the study of air transport networks that help to compare different airlines and draw new insights (\cite{Costa2018,Jiang2019,Mikko2014,Lordan2017}).
	
\begin{figure}
		\centering
		\includegraphics[width=0.6\textwidth]{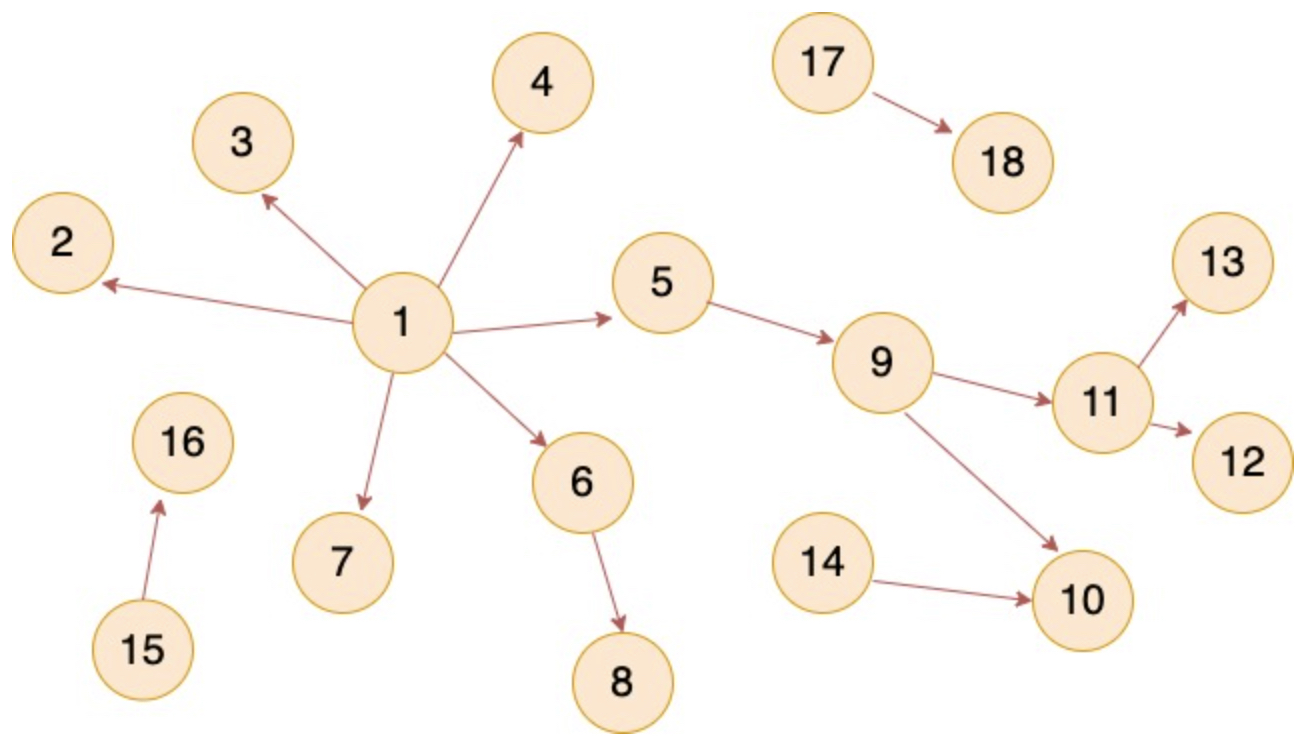}
		\caption{Representation of an arbitrary airline network: nodes 1 and 5, and edge (1,5) are the most disruptive elements in the network, and big disruptions could be avoided by avoiding delay to these disruptive elements.}
		\label{fig_idea}
	\end{figure}

	In this study, we propose a network science approach to identify the most disruptive elements in an airline flight network. Disruptive elements are those elements of the airline network that can cause a large disruption (according to the flight schedule of an airline) or have actually caused the biggest disruptions (according to historical operational data). Therefore, disruptive elements are the major causes (potential and actual) of delay propagation, and they can be used to improve airline operations. The proposed approach first models the airline schedule and historical operational data as a network, and then network science techniques are applied to identify disruptive elements. For example, suppose Fig.~\ref{fig_idea} models an airline’s schedules/delays (refer to Sec.~\ref{sec_network_science} for different ways of modeling airline data), where nodes refer to flights and edges refer to flight-connections. In this network, nodes 1 and 5 and edge (1,5) are among the most disruptive elements, as they are among the most influential elements in the network. Analysis of the airline schedules gives potential disruptive elements, whereas analysis of historical data, specifically delay propagation, gives actual disruptive elements. For example, Fig.~\ref{fig_idea} shows a delay network in which each node is a delayed flight and each edge is a delay propagation. From the figure, it is evident that the delay in flight 5 leads to delay propagation in flights 9, 10, 11, 12, and 13. Hence, there is a need to identify the disruptive elements of an airline network to improve operations because avoiding delays to disruptive elements’ can avoid large disruptions (\cite{brueckner2021airline}.
	
	The contributions of this study are summarized below.
	\begin{itemize}
		\item We present a model of an airline’s data (refer to Subsec.~\ref{subsec_types} for details) into airport--flight, flight--flight-connection, and multilayer flight--flight-connection networks, each with two variations as connection network (CN) and delay network (DN), modeling schedules and historical operational data, respectively. This modeling approach enables us to pinpoint the major causes of flight delay propagation, in terms of airports, flights, flight connections, and connection type.
		
		\item We propose the application of network science techniques, like centrality and percolation, to find the disruptive elements’ of an airline’s flight network. This work helps to identify influential/central airports, flights, flight connections, and connection type from an airline flight network, which can cause or have caused major disruptions. Modeling of schedules gives us potential disruptive elements whereas modeling of historical operational data provides us the actual disruptive elements. We also provide a ranking system to identify and prioritize the disruptive elements. Thus, the proposed approach helps find issues with the airline schedule and operations to identify disruptive elements that need prioritization to improve the operations of an airline.
		
		\item The proposed approach is validated with a case study of an airline; we observed that the potential disruptive elements of the airline have been identified as the actual disruptive elements. This indicates an issue with the airline schedule itself for causing disruptions and requires immediate attention to improve operations. The airline network also observes that small-world effect and delays can propagate to any part of the network in four flights. We also observe that passenger connections between flights are the most disruptive connection type.
		
		\item The analysis provides a tool to airlines to identify airports, flights, and flight connections that cause big disruptions. It also helps analyze the schedules and identify the issues with the schedule by looking for the overlap between potential disruptive elements (in CN) and actual disruptive elements (in DN). The findings of this analysis provide insights to an airline into how they can make their schedule robust to the disruptions by paying attention to the disruptive elements and making extra arrangements, such as increasing the slack time for disruptive flights and arranging alternate crew members for disruptive crew connections.
	\end{itemize}
	
	The remainder of the paper is organized as follows. Section~\ref{sec_literature} presents a literature review; Section~\ref{sec_network_science} describes the network science approach used and modeling of data into networks. A case study of an airline is discussed in Section~\ref{sec_analysis} and the concluding remarks and research implications are presented in Section~\ref{sec_conclusion}.
	
	\section{Literature review}
	\label{sec_literature}
	\subsection{Network science approach}
	\label{subsec_literature_net_sc}
	Network science has been widely used to analyze and provide new insights into real- life problems across different disciplines, from social science, engineering, technology, and biology, to airline problems (\cite{Newman2018}). In airline problems, data have been modeled as different types of networks, and the analysis has been targeted at drawing different types of insights. Airline problems can be categorized according to the type of data and the corresponding network modeling: single airline networks (\cite{Ledwoch2022,Reggiani2010}), multiple airline networks (\cite{Du2018,Jin2019,Yang2019}), airport networks (\cite{Baruah2019,Lordan2019,Song2017,Wang2019}), and multilayer airline network problems (\cite{Costa2018,Mikko2014,Lordan2017}).
	
	Airports are considered nodes in airport networks, and flights between airports are used as links/edges. For example, \cite{Lordan2019} discussed the global airport network, divided the network into seven regions, compared the network properties of different regions, and examined robustness by finding the core and critical cities. \cite{Baruah2019} compared airport network of India for three major airlines: Indigo, Jet Airways, and Air India, using network parameters and performed robustness analysis of airlines by studying the change of parameters on the removal of key nodes from the network. \cite{Wang2019} analyzed the resilience of an airport network by considering structural and dynamical aspects and by combining network science and operational dynamics. \cite{Olariaga2018} studied the configuration of traffic in the airport network caused by the liberalization of the air transport industry of Columbia . \cite{Baruah2019,Couto2015,cumelles2021cascading,Hossain2017,Jia2014,Jiang2017} analyzed the structural properties of the airport networks of China, Australia, U.S., Brazil, and India. \cite{Wandelt2019} presented a review and comparative analysis of evolution of domestic airport networks for India, U.S., Europe, China, Russia, Brazil, Australia, and Canada.
	
	Single- and multiple-airline problems focus on the analysis of single-airline data or comparative analysis of multiple airlines. For example, \cite{Reggiani2010} analyzed the Lufthansa airline network to understand the changing patterns in network configurations. \cite{Ledwoch2022} discussed a case study of an anonymous airline to determine the main  causes and effect of delay propagations through the airline’s flight network, as well as frequent delay patterns. \cite{Yang2019} discussed the structural properties of the Chinese airline network and \cite{Jin2019} used a network motif to analyze the structural characteristics of China’s passenger airline networks. \cite{Du2018} discussed the structure and dynamics of delay propagation using delay causality networks based on the Granger causality test. \cite{Alderighi2007} assessed the point-to-point and hub-and-spoke airline configurations based on spatial and temporal dimensions, and they concluded that temporal analysis is helpful to provide clear distinction between full-service carriers and low-cost carriers, whereas spatial analysis helped to find peculiarities within groups. \cite{Lordan2015} analyzed the robustness of three airline alliances (i.e., SkyTeam, oneworld, and Star alliance) through giant-component size vs. number of isolated airports and by using a multi-scale vulnerability measure; they concluded that Star alliance is the most resilient. \cite{Lordan2014} proposed three levels of analysis: airlines, airline alliances, and global route network, to analyze the robustness of air transport networks.
	
	Multilayer networks consider different types of relationships/connectivity between nodes and add additional value to the network analysis (\cite{Mikko2014}). For example, \cite{Lordan2017} analyzed the multilayered structure of European airlines using cores, bridges, and peripheries as layers of the network; they concluded that the network is more robust to the isolation of core nodes than the isolation of a combination of bridge and core nodes. \cite{Hong2016} studied the structural properties of the Chinese multilayered air transportation network using different airlines as layers, and major and low cost airlines. They established that the rich-club effect and small-world properties were mainly caused by major airlines. \cite{Costa2018} analyzed the multilayered and time varying structure of the Brazilian air transport network using different airlines as layers, and unveiled the strategies used by airlines to deal with the disruptions in the flight network and assessed the impact of the economic crisis on airlines. \cite{Jiang2017b} studied the transition point of air transport network using airlines as layers; they established that Chinese air transport has some scope to improve their operations while European airline network is already operating at its near-optimum level. \cite{Wang2017} analyzed the multilayered air transport network using airport, airlines, air route, and air traffic management networks as layers; they concluded that for flight delays, airports with similar geographical locations exhibit similar dynamics and delays or failure propagation decays slowly.
	
	Centrality is an important concept in network science that helps to identify influential nodes and edges in a network (\cite{Newman2018}). \cite{Clark2018} discussed the robustness and recovery strategies of U.S. airport network using network science techniques including different centrality measures. Additionally, \cite{Sathanur2019} studied U.S. airport network from the perspective of delay propagation and identified critical/influential airports based on influence maximization algorithm and diffusion simulator. \cite{Jin2019} studied the robustness of a network and identified critical nodes’ group in the network using eight different attack strategies, including random failure, target attack, and they were based on different centrality measures. \cite{Du2017} studied the Chinese air route network and identified the vital edges in the network using a memetic algorithm; they demonstrated that topologically important edges are not necessarily vital edges. \cite{Li2018} analyzed the Chinese transport network as a multilayered network and identified influential nodes based on evidence theory. Additionally, \cite{Cong2016} modeled the Chinese air transport network using airports as nodes and the correlations between air traffic flow of airports as edges; they identified the critical airports based on spatial and temporal correlations among airports.
	
	\subsection{Flight delay propagation}
	\label{subsec_literature_delayProp}
	Extensive research has been conducted on flight delay predictions, and various approaches have been used to study and predict delays using network science, statistical modeling, and machine learning. For example, \cite{Wang2003} developed an analytical model to study delay propagation in U.S. airports that separates the fixed and variable parts of delays and delay propagation. \cite{Abdelghany2004} modeled the airlines data into directed acyclic graph  and used the shortest path algorithm to predict flight delay propagations for irregular operations. \cite{Jani2005} developed an analytical model for hub-and-spoke airline network to estimate the economic impact of large disruptions. \cite{Xu2005} developed a Bayesian network to investigate and visualize delay propagation among airports. Additionally, \cite{Liu2008} used the Bayesian network to model the arrival delays at a hub airport and study the delay propagation within and from the airport. \cite{AhmadBeygi2008} proposed propagation trees to study the potential of delay propagation from the airline schedules and their actual operations. \cite{Wong2012} used the Cox proportional hazards model to study delay propagation in Taiwanese domestic airlines. Recently, many machine learning-based models (e.g., deep learning (\cite{Gui2020}), regression (\cite{Gopalakrishnan2017}), classification and regression trees (\cite{Gopalakrishnan2017}), support vector machines (\cite{Wu2019}), gradient boosting classifiers (\cite{Thiagarajan2017}), and random forests (\cite{Gopalakrishnan2017})) have been used to predict future flight delays.
	
	In this study, we modeled data from a single airline in different connections, delays, and multilayer networks. Additionally, we utilized network science techniques, such as centrality and percolation, to identify disruptive airports, flight connections, and flights that can affect several other flights (according to the airline schedule) or that actually lead to the highest delay propagations to other flights (according to the historical operational data). This study is a novel application of network science techniques and provides a useful tool for airlines to improve their flight operations by focusing on the most disruptive elements and building robust schedules. This study addresses the problem of delay propagation; however, it differs from the abovementioned literature in terms of modeling data into networks, analysis, and the benefits of the analysis. To the best of our knowledge, this is the first study focusing on the major causes of delay propagation. Similar to \cite{AhmadBeygi2008}, the authors analyzed schedules and operational data for delay propagation; however, they used propagation trees to study the potential of delay propagation using a network science approach to analyze schedules and operational data to identify the major causes of delay propagation.

	\section{Network science approach}
	\label{sec_network_science}
	Here, we present the modeling of ’the schedules and historical operational data of an airline, as well as different network science properties and techniques for studying the network of the airline’.
	
	\subsection{Types of networks}
	\label{subsec_types}
	Network modeling provides insights that can be drawn using the network science approach. Therefore, we have modeled the problem mainly using three types of networks: connection, delay, and multilayer flight connection networks, which are discussed below. Their summary is presented in Table~\ref{tab:networks}.
	
	\begin{table}[htb]
		\caption{Summary of different network modeling approaches.}
		\label{tab:networks}
		\begin{tabular}{ p{1.1cm}p{7cm}p{7cm}}
			\hline
			\multicolumn{1}{c}{\textbf{Network}} & \multicolumn{1}{c}{\textbf{Description}} & \multicolumn{1}{c}{\textbf{Insights}} \\ \hline
			\textbf{AFCN} & Node: airports, Edge: two airports have an edge between them if there is at least one flight between them, Edge weight: number of flights between the airports. & AFCN helps to analyze the airline schedules. The analysis of AFCN yields the busiest airports and airport pairs, which have highest traffic between them. \\ 
			\textbf{AFDN} & Similar to AFCN but here we consider only delayed flights. & AFDN helps to analyze the airline historical operational data. It helps to identify the airports and airport pairs which caused the biggest disruptions. \\ 
			\textbf{FCCN} & Node: flights, Edge: shared flight connection, Weight: frequency of connection & FCCN helps to analyze the airline schedules. It gives us flights and flight connections, which have potential to cause disruption to several others. \\
			\textbf{FCDN} & It is similar to FCCN except that it considers only delayed flights as nodes and flight connections with delay propagation only. & FCDN helps to analyze the airline historical operational data. It helps to identify flights and flight connections, which caused the biggest disruptions in the network. \\
			\textbf{MLCN} & Multilayer Network adds extra value to the network analysis by considering the types of connectivity in the network. MLCN are similar to FCCN except that each layer considers connections of one type. & MLCN helps to analyze the airline schedules. It helps to identify the connection type which can cause the biggest disruptions in the network. \\ 
			\textbf{MLDN} & MLDN are similar to FCDN except each layer considers connections of one type. & MLDN helps to analyze the airline historical operational data. It helps to identify the connection type, which caused the biggest disruptions in the network. \\ \hline
			\multicolumn{3}{l}{%
				\begin{minipage}{15cm}%
					\footnotesize Note: AFCN: Airport-Flight Connection Network, AFDN: Airport-Flight Delay Network, FCCN:~Flight-Connection Connection Network, FCDN: Flight-Connection Delay Network, MLCN: Multilayer Connection Network, MLDN:~Multilayer Delay Network
				\end{minipage}%
			}\\	
		\end{tabular}
	\end{table}

	\subsubsection{Connection network}
	\label{subsub_con_net}
\begin{figure*}[htb!]
		\centering
		\begin{subfigure}[t]{0.33\textwidth}			
			\centering
			\includegraphics[width=\textwidth]{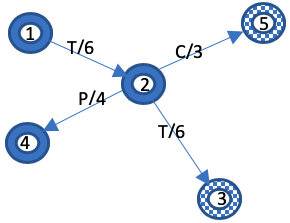}
			\caption{}
			\label{subfig_con_net}
		\end{subfigure}%
		~ 
		\begin{subfigure}[t]{0.3\textwidth}			
			\centering
			\includegraphics[width=\textwidth]{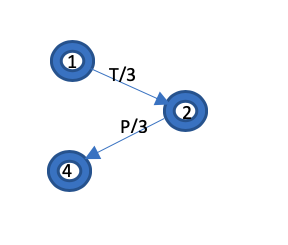}
			\caption{}
			\label{subfig_del_net}
		\end{subfigure}
		~ 
		\begin{subfigure}[t]{0.3\textwidth}			
			\centering
			\includegraphics[width=\textwidth]{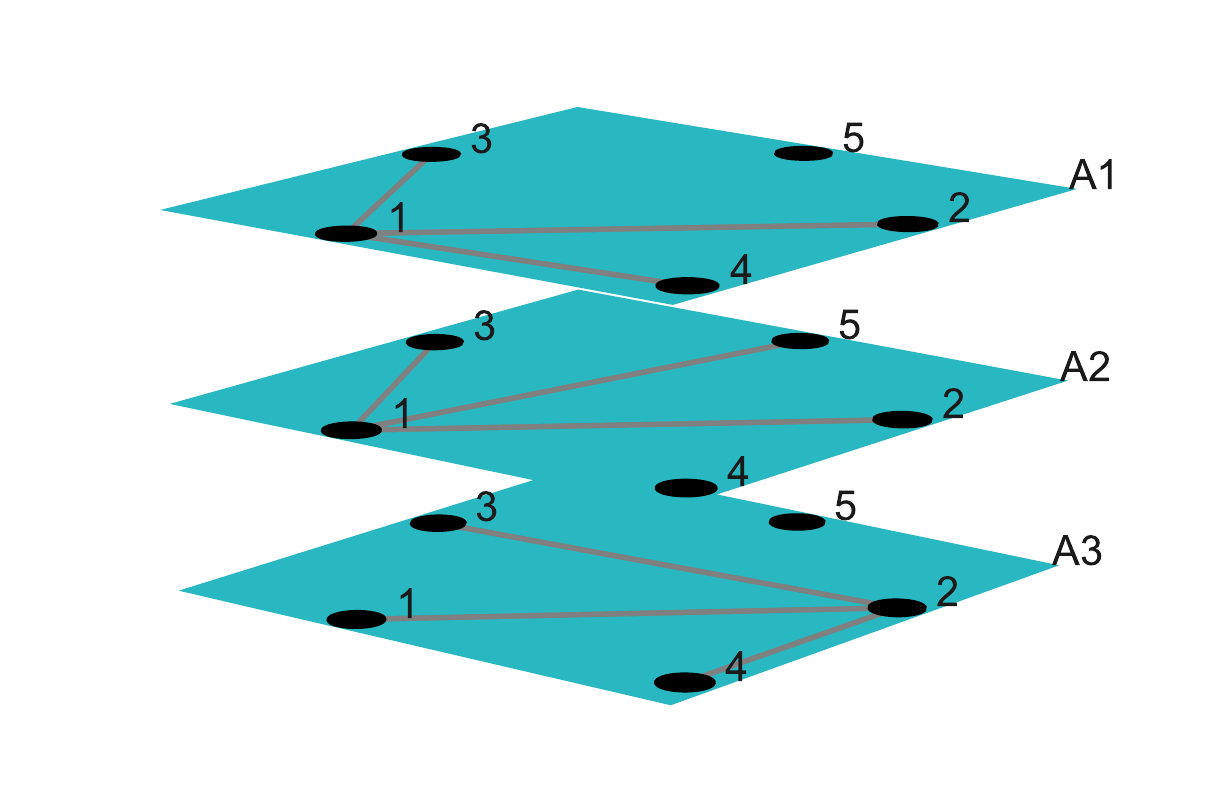}
			\caption{}
			\label{subfig_mln}
		\end{subfigure}		
		\caption{Network types: solid rings are the delayed flights and others are on time flights; T, C, and P refer to tail, crew, and passenger connections, respectively, and number refers to the frequency of flights. (a) Flight-connection connection network (FCCN); (b) Flight-connection delay network; (c) An example of multilayer network with 5 nodes and three layers, viz., A1, A2, and A3.}
		\label{fig_networs}
	\end{figure*}

	A CN models the schedules of an airline along with information about possible shared connections in terms of passengers, crew, and tail (i.e., aircraft) between connecting flights. It considers the scheduled arrival, departure, origin, and destination airports; however, it does not consider the actual operations of the data, such as the actual arrival and departure of flights. CNs cover the schedule for a given period. There are three types of CNs: airport-flight connection network (AFCN), flight-connection connection network (FCCN), and multilayer connection network (MLCN), as discussed below (MLCN is discussed later in \ref{subsubsec_multi_net}). An example of FCCN is presented in Fig.~\ref{fig_networs}(a). CNs help analyze the airline schedule and identify potential disruptive elements as the analysis is on schedule. They help to identify the busiest elements of the network, which can cause large disruptions. We can crosscheck the disruptive elements from a CN with the corresponding delay network (DN; discussed below), and is the lack of overlap between the disruptive elements in CN (potential disruptive elements) and in DN (actual disruptive elements), this indicates that the airline schedule has no major issue. Otherwise, overlap indicates that something might be wrong with the schedules, and the airline should fix the overlapped disruptive elements first.
	
	\textbf{AFCN:} In AFCN, airports are the nodes and flights are considered as edges with direction from origin airport to destination airport. Two airports have an edge between them if there is at least one flight between them; if there is more than one flight between the airports, this is represented as the weight of the edge. The analysis of AFCN gives us potential disruptive airports and airport pairs, as discussed in \ref{subsec_connection}, which have the highest traffic between them.
	
	\textbf{FCCN:} Each flight is taken as a node, where flight is represented using flight number, scheduled departure time, origin airport, and destination airport. For instance, suppose flight number 100 is scheduled to depart at 10:10 from airport A1 to airport A2, then 100.10:10.A1.A2 are considered unique flights and nodes, respectively. If two flights share passengers, crew, or tail, this is called a flight connection and is considered as the edge of the network with direction from the first flight to the second flight. The frequency of flight connections between flights is considered as the weight of the edges. For example, suppose flight number 100 flies every day, its weight would be seven in one week. The analysis of FCCN gives us potentially disruptive flights and flight connections, as discussed in \ref{subsec_FCN}, which are flights that have the potential to disrupt several other flights.
	
	\subsubsection{Delay network}
	\label{subsubsec_del_net}
	A DN models the historical operational data of an airline, focusing mainly on delays and their propagation through the network. DNs can be seen as a subset of CNs considering only delayed flights and flight connections, which lead to delay propagation. For example, Fig.~\ref{fig_networs}(b) represents the flight-connection DN. DNs help analyze disruptive elements from historical data, as they consider actual departure, arrival times, and delays propagated when the airline schedule is executed. This helps identify the flights and flight connections that lead to the largest disruptions in the network. DNs can be used to review airline operations and make decisions to improve operations by focusing on the disruptive elements in the network. There are three types of DNs: airport flight delay networks (AFDNs), light-connection delay networks (FCDNs), and multilayer delay networks (MLDNs). MLDN is discussed in Section \ref{subsubsec_multi_net}.
	
	\textbf{AFDN:} It is similar to the AFCN, except that AFDN considers only delayed flights for forming edges and delayed flight frequencies as weight of the edge. AFDN helps identify disruptive airports and airport pairs that actually caused the biggest disruptions, as discussed in \ref{subsec_connection}.
	
	\textbf{FCDN:} It is similar to the FCCN, except that it considers only delayed flights as nodes and flight connections with delay propagation. The characterization of the delay propagation is as used in \cite{Ledwoch2022} and the propagated delay is equal to the arrival delay of the incoming first flight minus the slack time. The slack time is the time difference between the departure of the second flight (connecting flight) and the arrival of the first flight minus the minimum ground time to prepare for the second flight. The delay propagation frequency of the flight connection is used as the weight of the edge. FCDN helps identify disruptive flights and disruptive flight connections that actually cause the biggest disruptions in the network, as discussed in \ref{subsec_FCN}.

	\subsubsection{Multilayer flight connection network}
	\label{subsubsec_multi_net}
	Multilayer networks (ML) are networks where interactions between elements of the network may change over time, have different types of connections, or include other types of complications; thus, they can be divided into subsystems depending upon these interactions (for details refer to \cite{Mikko2014}). These networks add extra value to the network analysis by considering the types of connectivity in the network. We used three types of connections: crew, tail, and passenger connections between flights as the different layers. There are two types of MLs: MLCN and MLDN . MLCN and MLDN are similar to FCCN and FCDN, respectively, where each layer is an FCCN/FCDN network with one-type connections. An example of ML is shown in Fig.~\ref{fig_networs}(c), which has five nodes common to all layers, and three layers with their own edges depending on the type of edge.
	\begin{table}[htb!]
		\caption{Network science techniques/properties and insights drawn from them.}
		\label{tab_properties}
		\begin{tabular}{ p{3cm}p{7cm}p{6cm}}
			\hline
			\multicolumn{1}{c}{\textbf{Property}} & \multicolumn{1}{c}{\textbf{Definition}} & \multicolumn{1}{c}{\textbf{Insights}} \\ \hline
			\textbf{Network} & Network is a set of nodes and set of links where nodes represent some objects and links capture some type of relationship between those objects. & It is used to represent and analyze data with non-linear relationships.  \\
			\textbf{Degree distribution} & Degree distribution represents the probability distribution of the degree over the entire network. & It provides us the structural insights about the network.  \\
			\textbf{Centrality} & It helps to identify the central elements of the network depending on their impact on other elements. &  This concept helps to identify the most influential flights, airports, and connections that have a significant impact on other flights, in terms of delay propagation.\\
			\textbf{Out degree Centrality} & It helps to identify the important nodes based on their out degree. &  This concept helps to identify the flights and airports that have a significant impact on other flights.\\
			\textbf{Node Betweenness Centrality} &  It helps to identify important nodes based on flow of information through them. &  This concept helps to identify influential flights and airports through which highest delays are either propagated or can potentially be propagated.\\ 
			\textbf{Edge Betweenness Centrality} & It  helps to identify important edges based on flow of information through them. &  This concept helps to identify flight connections and airport-pairs through which highest delays are either propagated or can potentially be propagated.\\
			\textbf{Percolation} &It is a study of removal of nodes/edges of a network and their effect on the connectivity of the network. & In the airline scenario, when applied to a delay network for node percolation, it tells us which flights delay should be avoided to avoid big disruptions.   \\ 
			\textbf{Diameter} & The diameter of a connected network is defined as the largest shortest distance between any two nodes and the diameter of a disconnected network is infinite.  & This concept tell us about the minimum number of flights required for the propagation of delays from any part of the network to any other part.  \\
			\textbf{Small-world effect} & When one can reach from any node of the network to any other node in the network in a few steps, then that network is said to exhibit the small-world effect.  & In an airline network, if small-world effect occurs in the network, it is susceptible to big disruptions because delays can propagate to the entire network with a few flight delays.  \\
			\textbf{Density} & It is defined as the ratio of the number of links in the network to the total number of possible links in the network. &  This is a good metric to compare the different layers of the MFCN. \\ \hline
		\end{tabular}
	\end{table}
	
	\subsection{Network science techniques and insights}
	\label{subsec_properties}
	Here, we discuss the different network properties and techniques used to analyze the airline network modeled in the previous subsection, along with the insights provided by them regarding the problem; the summary is presented in Table~\ref{tab_properties}.
	
	\textbf{Network:} Networks are used to represent data with non-linear relationships and can be defined as a set of nodes $V$ and a set of links $E$, where nodes represent some objects and links capture the relationship between those objects. For example, in an airport-flight network, airports are the nodes and two nodes have a link between them if flights operate between them. Networks can be of two main types: directed and undirected. The network is directed when there is a sense of direction associated with links in the network; otherwise, it is called undirected. For example, flights take off from the origin airport to the destination airport; thus, the airport-flight network is a directed network.
	
	\textbf{Degree Distribution:} In a network, the degree of a node is the number of other nodes to which it is connected in an undirected network. For a directed network, we have an in-degree and out-degree, where the in-degree of a node is the number of nodes with incoming connections, and the out-degree of a node is the number of outgoing connections from the node to other nodes. The degree distribution represents the probability distribution of the degree over the entire network, and it is useful for providing structural insights into the network.
	
	\textbf{Centrality:} As the name suggests, it helps to identify the central elements of the network depending on their impact on other elements. Thus, this concept helps identify flights, airports, and connections that significantly affect other flights in terms of delay propagation. Here, we study out-degree centrality and betweenness centrality. The first helps to identify the important nodes based on their out-degree, that is, flights that lead to delayed propagation to others. The second helps to identify important nodes and edges based on the flow of information through them; that is, to identify flights and connections through which delays propagate. The betweenness centrality of node $v$ is given by:
	\begin{equation}
		c_B \left( v \right) = \sum_{s,t\in V} \frac{\sigma\left(s,t|v\right)}{\sigma\left(s,t\right)},
	\end{equation}
	where $V$ is the set of vertices, $\sigma\left(s,t \right)$ is the number of shortest paths between all pairs of nodes $\left(s,t\right)$, and $\sigma\left(s,t|v\right)$ is the number of shortest paths between all pairs $\left(s,t\right)$ through node $v$. The edge betweenness centrality of edge $e$ is as follows:
	\begin{equation}
		c_B \left( e \right) = \sum_{s,t\in V} \frac{\sigma\left(s,t|e\right)}{\sigma\left(s,t\right)},
	\end{equation}
	where $\sigma\left(s,t|e\right)$ denotes the number of shortest paths between all pairs of nodes $\left(s,t\right)$ through edge $e$. The out-degree centrality of node $v$ is expressed as
	\begin{equation}
		c_D \left( v \right) =  \frac{o(v)}{n},
	\end{equation}
	where $o(v)$ is the out-degree of node $v$ and $n$ is the number of nodes. Generally, degree centrality is normalized by dividing the value by the maximum possible value, that is, $n-1$.
	
	\textbf{Percolation:} It is the study of removal of nodes/edges of a network and their effect on the connectivity of the network. In an airline scenario, when applied to a delay network for node percolation, flight delays should be avoided to avoid significant disruptions. It can be observed as dynamic betweenness, where we calculate betweenness of the network and remove the node/edge with the highest value, then betweenness is calculated for the network and then the node/edge with the highest value is removed, and so on.
	
	\textbf{Diameter:} Diameter of a connected network is defined as the largest shortest distance between any two nodes. The diameter of a disconnected network is infinite because some nodes cannot be reached. This concept indicates the minimum number of flights required to propagate delays from any part of the network to any other. The smaller the diameter, the better the air connectivity; however, the greater are the chances of disruptions because fewer flights would be required for the propagation of delays to the entire network.
	
	\textbf{Small-world effect:} When one can reach from any node of the network to any other node in the network in few steps, then the network is said to exhibit small-world effect. In an airline network, if the small-world effect occurs, the network is susceptible to large disruptions because delays can propagate to the entire network with just a few flight delays. This effect can be measured using the diameter of the network; therefore, if the network has a small diameter, it exhibits a small-world effect.
	
	
	%
	
	\textbf{Density:} It can be defined as the ratio of the number of links in the network to the total number of possible links in the network. This can be a good metric for comparing different layers of a MFCN. The density of a network is given as
	\begin{equation}
		d = \frac{m}{n(n-1)},
	\end{equation}
	where \(m\) denotes the number of edges in the network. The density of the network is zero if it has no edges and one if it is a complete network.

	\section{Case study}
	\label{sec_analysis}
	Here, we present a case study of an operating airline to demonstrate the proposed methodology. We discuss the data used in the analysis and the results of the analysis with different types of networks to which data has been modeled, namely, airport flight, flight connection, and MFCNs. Each network has two variations: CN and DN, as discussed in Sec. ~\ref{sec_network_science}, and we used the out-degrees to discuss the results because the in-degrees are similar to the out-degrees for airline networks and they can be used to track the propagation of disruptions/delays from one airport/flight to another. Analysis of the CNs revealed that the potential disruptive elements and DNs reveal the actual disruptive elements of the network. Many cases arise from the results of these two analyses, which should be interpreted carefully. Therefore, we have developed the following ranking system (in order of descending priority) to prioritize fixing the disruptive elements of an airline for developing robust operations.
	\begin{enumerate}
		\item[(1)] If potential and actual disruptive elements overlap and the overlapped elements come up across multiple techniques, such as disruptive elements using percolation and disruptive elements using betweenness centrality, the disruptive elements are considered to have the highest priority and should be addressed first, to improve the airline operations. This is because these elements indicate an issue with the airline schedule.
		
		\item[(2)] If potential and actual disruptive elements overlap and the overlapped elements occur only once, it indicates an issue with the airline schedule.
		
		\item[(3)] If there is no overlap between potential and actual disruptive elements; however, some of the \textit{actual} disruptive elements come up in multiple techniques, the disruptive elements need attention as they have caused disruptions in the operations.
		
		\item[(4)] If there is no overlap between potential and actual disruptive elements, and no element comes in multiple techniques,  there is no major issue with the airline schedule. It is not possible to prioritize the disruptive elements in this case; however, the analysis is still useful for an expert, for identifying disruptive elements that can be prioritized.
		
		\item[(5)] If there is no overlap between potential and actual disruptive elements, and some of the \textit{potential} disruptive elements occur in multiple techniques, the disruptive elements only identify the busiest elements of the network, and from the current analysis they appear to be well taken care of and do not require more attention.
		
	\end{enumerate}

	\subsection{Data}
	\label{subsec_data}
	We conducted a case study of an airline with a hub-and-spoke topology (a method of connecting nodes in a network with most of the nodes connected to one central node and other nodes having few connections among them), as confirmed by the results, whose name is not revealed on request of the airline. The data received from the airline for a period of six months include flight details (i.e., aircraft used, origin and destination airports, scheduled and actual departure and arrival times, minimum ground times) few anonymized details about passengers (i.e., the number of passengers of different travelling classes) and crew connection details (i.e., crew ids) shared between connecting flights, as summarized in Table~\ref{table_data}. Flight delays are calculated as the difference between the scheduled and actual times of the flight, and delays are said to be propagated from one flight to another if the first flight is delayed and leads to the delay of the second flight. This is equal to the arrival delay of the incoming first flight minus the slack time, where the slack time is the time difference between the departure of the second flight (connecting flight) and the arrival of the first flight minus the minimum ground time to prepare for the second flight (for mathematical derivation, please refer to \cite{Ledwoch2022}. The data were anonymized and all flight numbers and airport names were assigned names 1, 2, 3,... and A1, A2, A3,..., respectively.
	
	\begin{table}[htb]
		\caption{Airline data description.}
		\label{table_data}
		\begin{tabular}{ll} \hline
			\textbf{Name} & \textbf{Description} \\ \hline
			Flight Number & This is number assigned to each flight and it might not be unique. \\
			Origin Airport & This is the airport from where the flight takes off. \\
			Departure Airport & This is the airport where the flight lands. \\
			STA & This is the scheduled time of arrival of the flight. \\
			ATA & This is the actual time of arrival of the flight. \\
			STD & This is the scheduled time of departure of the flight. \\
			ATD & This is the actual time of departure of the flight. \\
			MGT & This is the minimum amount of time a flight needs to get ready for the next flight. \\
			Passengers & Number of passengers travelling in different classes for the connecting flight. \\
			Crew ID & This is the unique (anonymized) id associated with each crew member. \\
			Crew flights & F lights served by a particular crew members.\\ \hline
		\end{tabular}
	\end{table}

	\subsection{Airport-flight network}
	\label{subsec_connection}
	\begin{figure*}[htb]
		\centering
		\begin{subfigure}[t]{0.48\textwidth}			
			\centering
			\includegraphics[width=\linewidth]{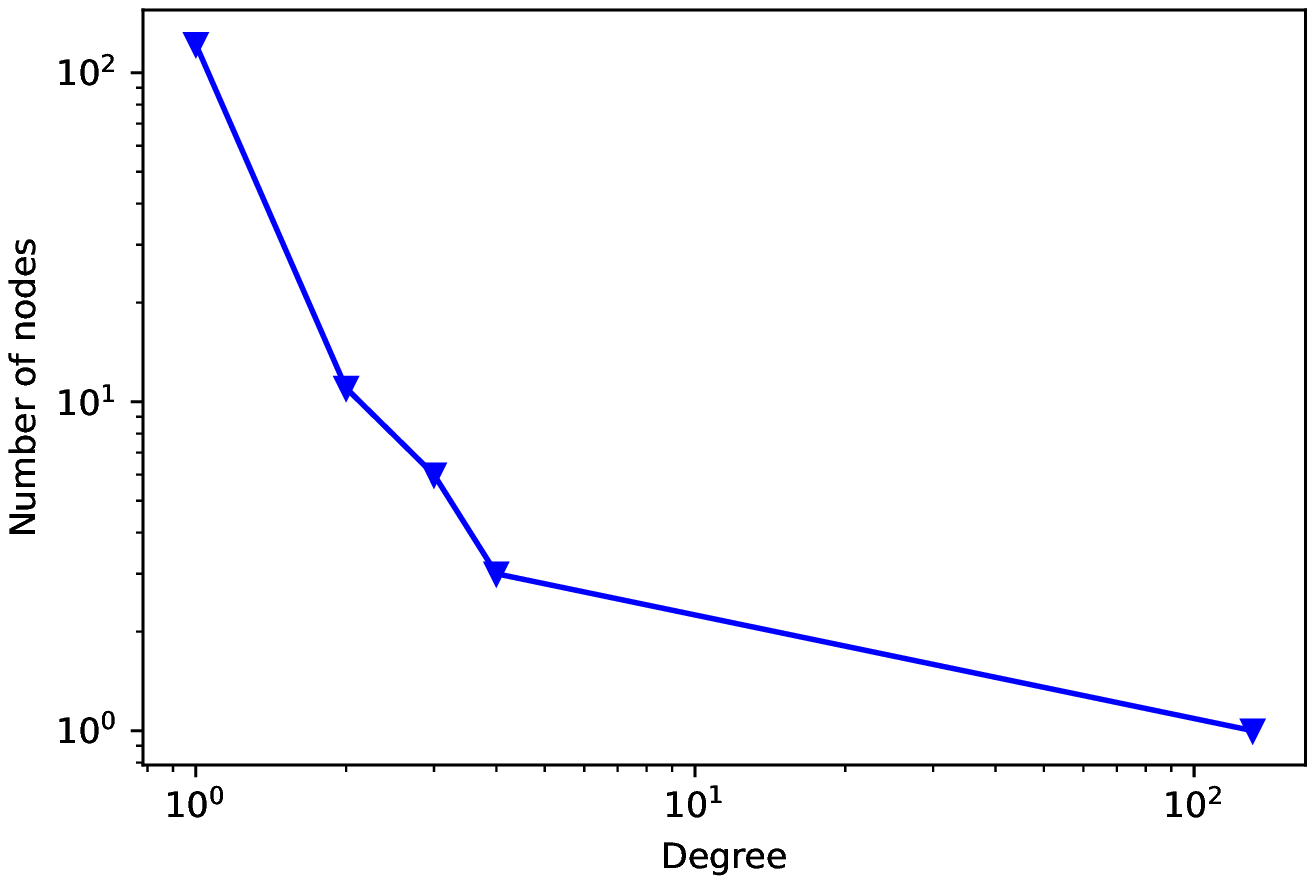}
			\caption{}
			\label{subfig_AF_CN}
		\end{subfigure}
		~
		\begin{subfigure}[t]{0.48\textwidth}			
			\centering
			\includegraphics[width=\linewidth]{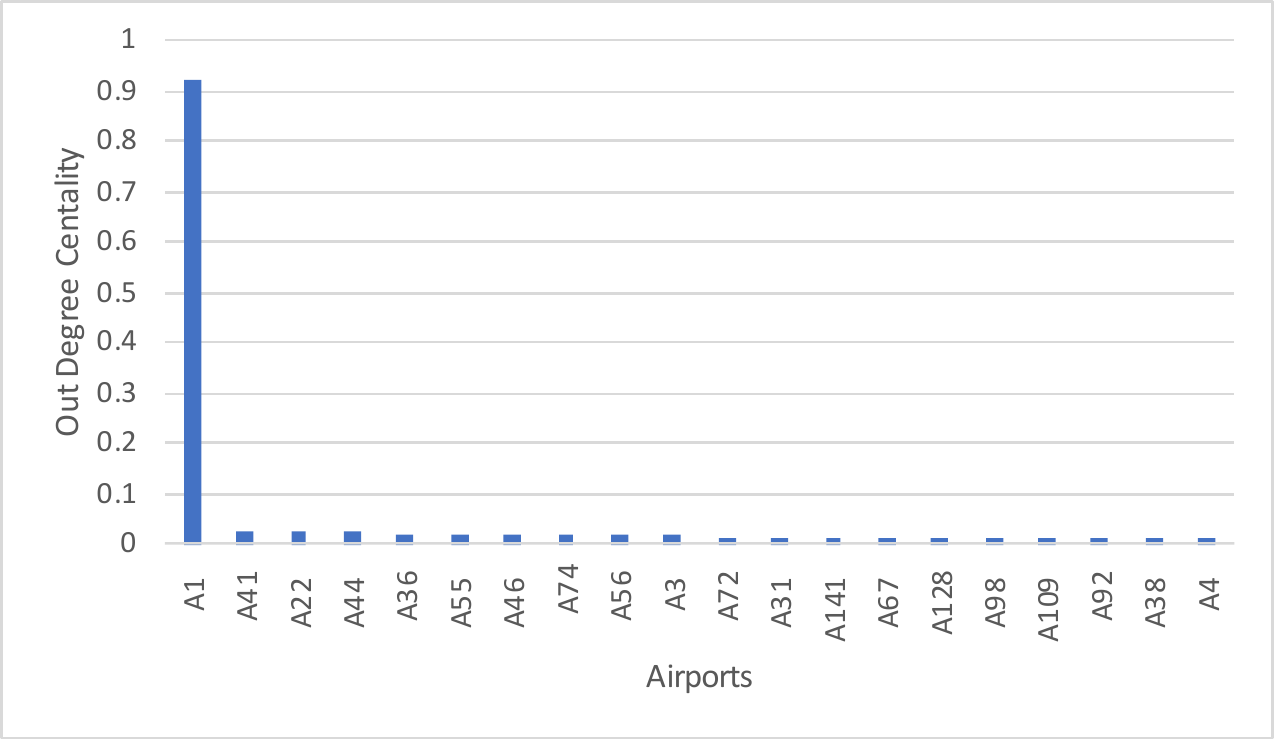}
			\caption{}
			\label{subfig_BA_D_AFN}
		\end{subfigure}%
		
		\begin{subfigure}[t]{0.48\textwidth}			
			\centering
			\includegraphics[width=\linewidth]{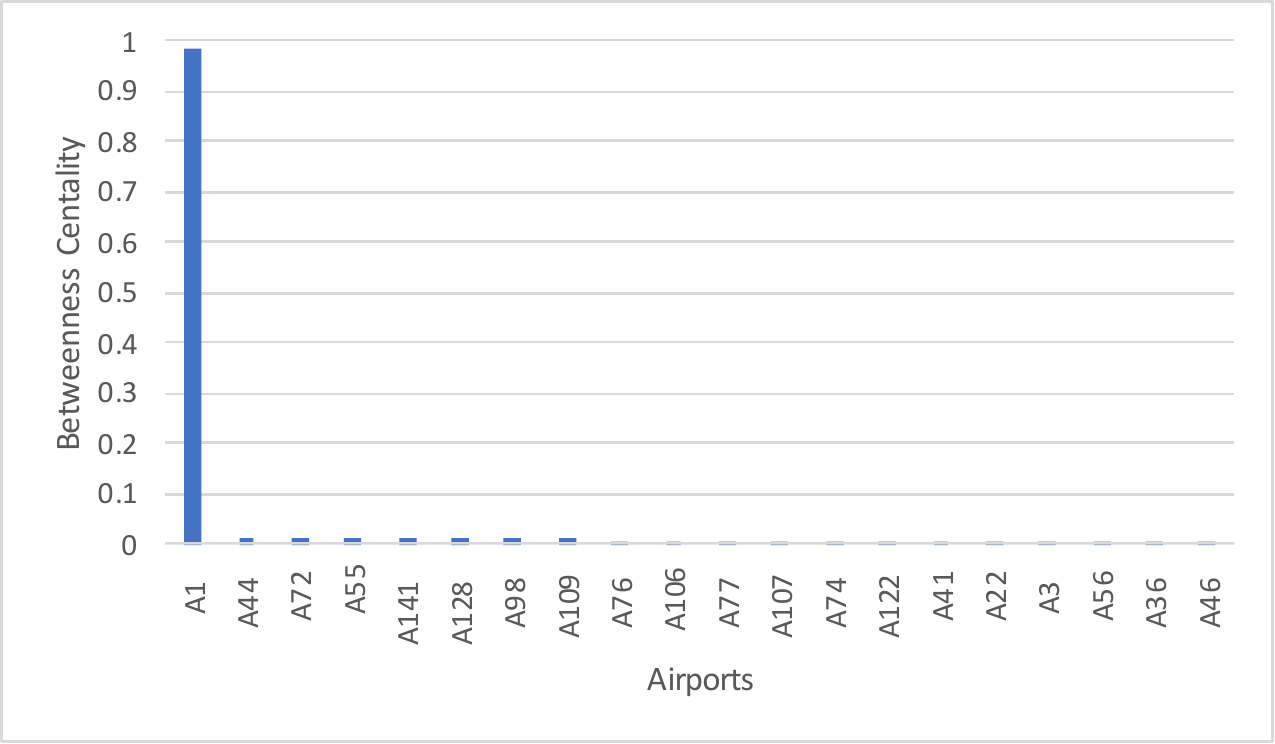}
			\caption{}
			\label{subfig_BA_B_AF}
		\end{subfigure}%
		~
		\begin{subfigure}[t]{0.48\textwidth}
			\centering
			\includegraphics[width=\linewidth]{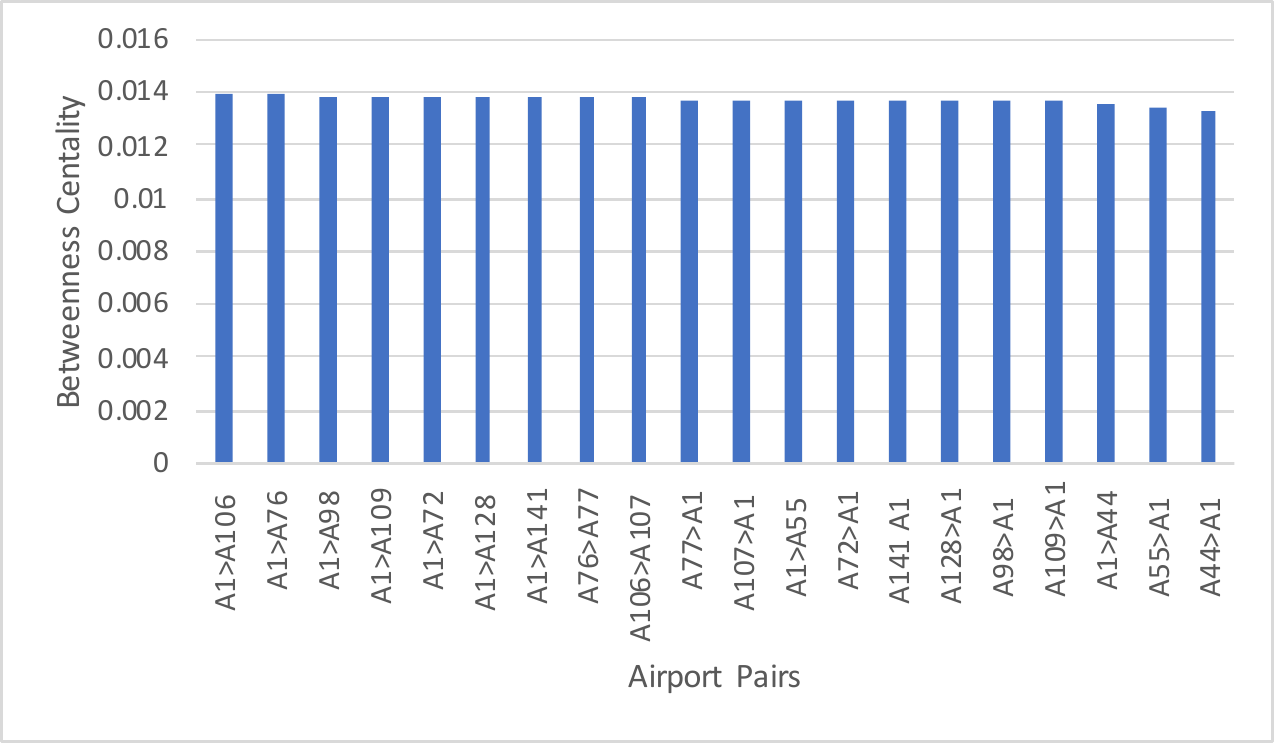}
			\caption{}
			\label{subfig_BF_B_AF}
		\end{subfigure}
		\caption{Analysis of airport-flight network. (a) Airport-flight network out-degree distributions; (b) Disruptive airports using out degree centrality in airport-flight network (AFN); (c) Disruptive airports using betweenness centrality in AFN; (d) Disruptive airport pairs using betweenness centrality in AFN.}
		\label{fig_AFN}
	\end{figure*}
	
	The diameter of an AFCN is four, which means that it will take a minimum of four flights to propagate a delay to the entire network of the airline. That is, the AFCN exhibits a small-world effect, which means that one node can reach any other node of the network with few edges (\cite{Newman2018}) . The results of AFN are presented in Fig. ~\ref{fig_AFN}. Disruptive elements in the CN are similar to those in the DN of the AFN. Unlike the flight-connection networks (FCNs) (shown in the next subsection), for AFN we have shown the results using a single figure for each category . This is mainly because AFN represents a high-level view of an airline, where it studies disruptive airports and disruptive airport pairs over a period of six months and does not go to flight level.
	
	Fig.~\ref{fig_AFN}(a) shows the out-degree distribution of the network. It is the same for the CN and the DN because if some airports have outgoing flights to four other airports, the DN will have connections to those airports because there is a high probability that at least one flight to the four airports will be late in six months. As shown in the figure, the degree distribution is quite different from that of real-life networks, which follow the power law (\cite{Cong2016}). This is because the network has one airport with a large number of connections to other airports, and most of the airports are connected to a few other airports, meaning that the network topology is hub-and-spoke, and there is one hub airport through which other airports are connected.
	
	Fig.~\ref{fig_AFN}(b) represents disruptive airports using out-degree centrality. From the figure, it is evident  that A1 is the central airport, that is, many connecting flights take off from A1, and there is a huge difference in terms of air traffic with other airports. Thus, A1 is the hub airport, and the network follows a hub-and-spoke topology, as already indicated by the degree distribution. Fig.~\ref{fig_AFN}(c) represents disruptive airports according to betweenness centrality and presents the top airports with high traffic flow and delay propagation through them. Once again, A1 airport has the highest traffic and delays because most flights pass through it. Fig.~\ref{fig_AFN}(d) represents disruptive airport pairs with the highest traffic flow and delay propagation. Most airports include A1 as an airport because A1 is the hub airport. Moreover, the top three disruptive airport pairs have flights from A1 to A106, A76, and A98; however, the difference was not significant between different pairs.
	
	Clearly, the hub airport A1 has the highest traffic and causes the highest number of delay propagations; thus, it needs the most attention to improve the airline’s operations. Moreover, there are overlaps among the results represented by Fig.~\ref{fig_AFN}(b)--(d); for example, A98 and A55 are common to all of them; therefore, they need the second-highest attention to improve the airline’s operations.

	\subsection{Flight-connection network}
	\label{subsec_FCN}
\begin{figure}[htb!]
		\centering
		\includegraphics[width=0.5\linewidth]{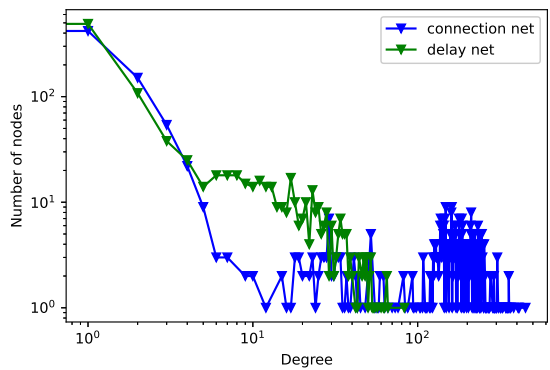}
		\caption{FCN out-degree distributions}
		\label{subfig_FC}
	\end{figure}

\begin{figure*}[htb!]
		\centering
		\begin{subfigure}[t]{0.5\textwidth}
			\centering
			\includegraphics[width=\linewidth]{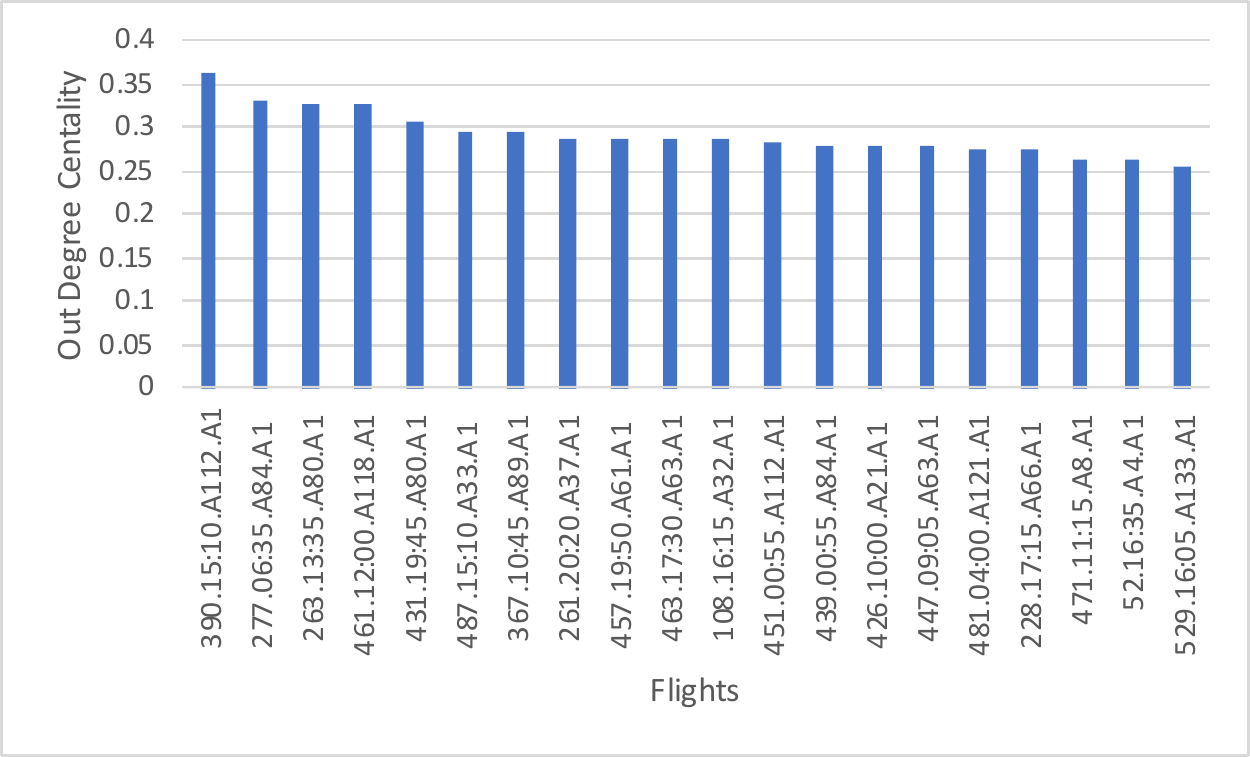}
			\caption{}
			\label{subfig_BF_D_FCN}
		\end{subfigure}%
		~ 
		\begin{subfigure}[t]{0.5\textwidth}
			\centering
			\includegraphics[width=\linewidth]{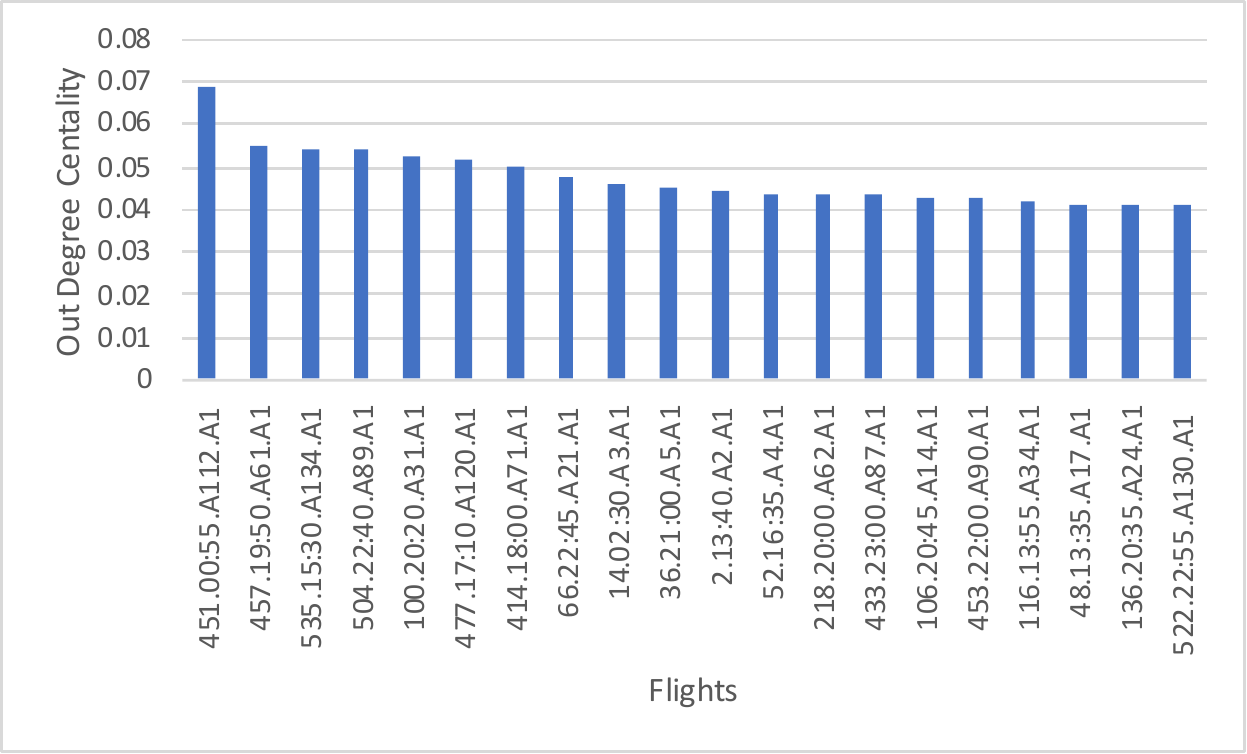}
			\caption{}
			\label{subfig_BF_D_FCDN}
		\end{subfigure}
		\caption{Disruptive flights using out degree centrality in FCN. (a) Connection network; (b) Delay network.}
		\label{fig_BF_D_FC}
	\end{figure*}

\begin{figure*}[htb!]
		\centering
		\begin{subfigure}[t]{0.5\textwidth}
			\centering
			\includegraphics[width=\linewidth]{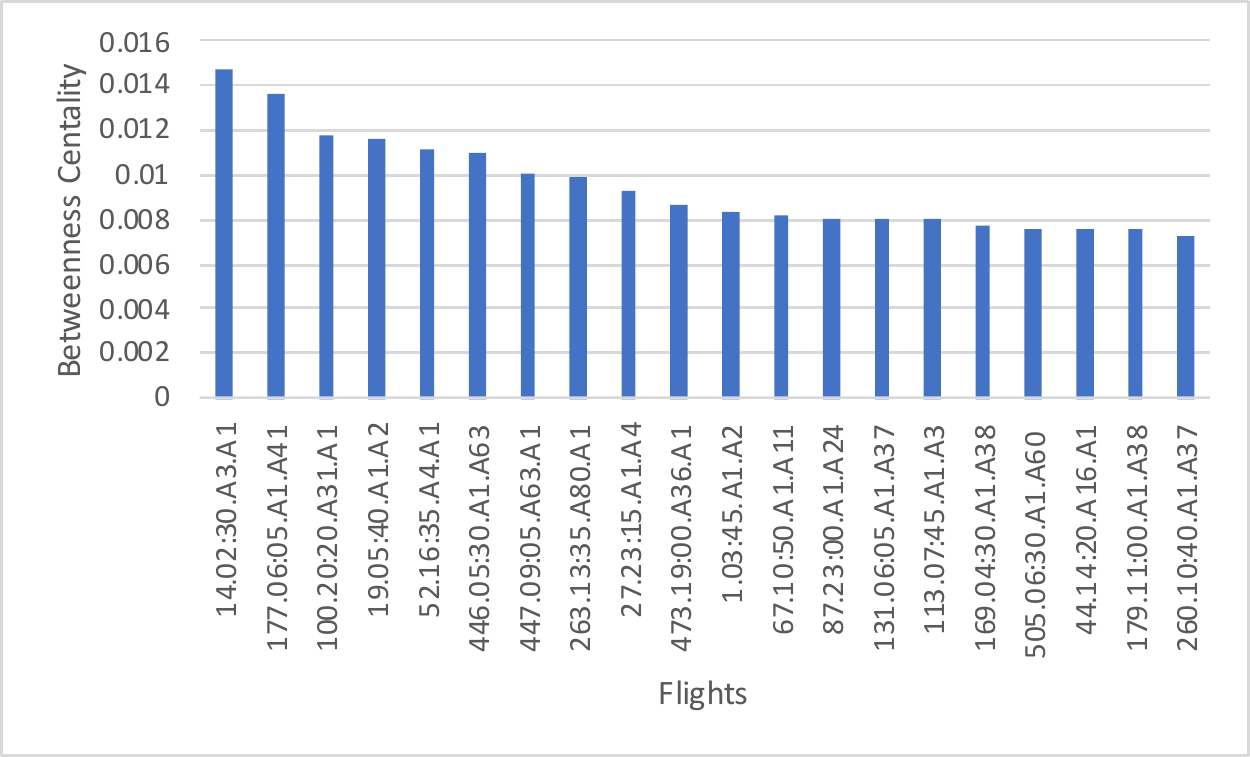}
			\caption{}
			\label{subfig_BF_B_FCN}
		\end{subfigure}%
		~ 
		\begin{subfigure}[t]{0.5\textwidth}
			\centering
			\includegraphics[width=\linewidth]{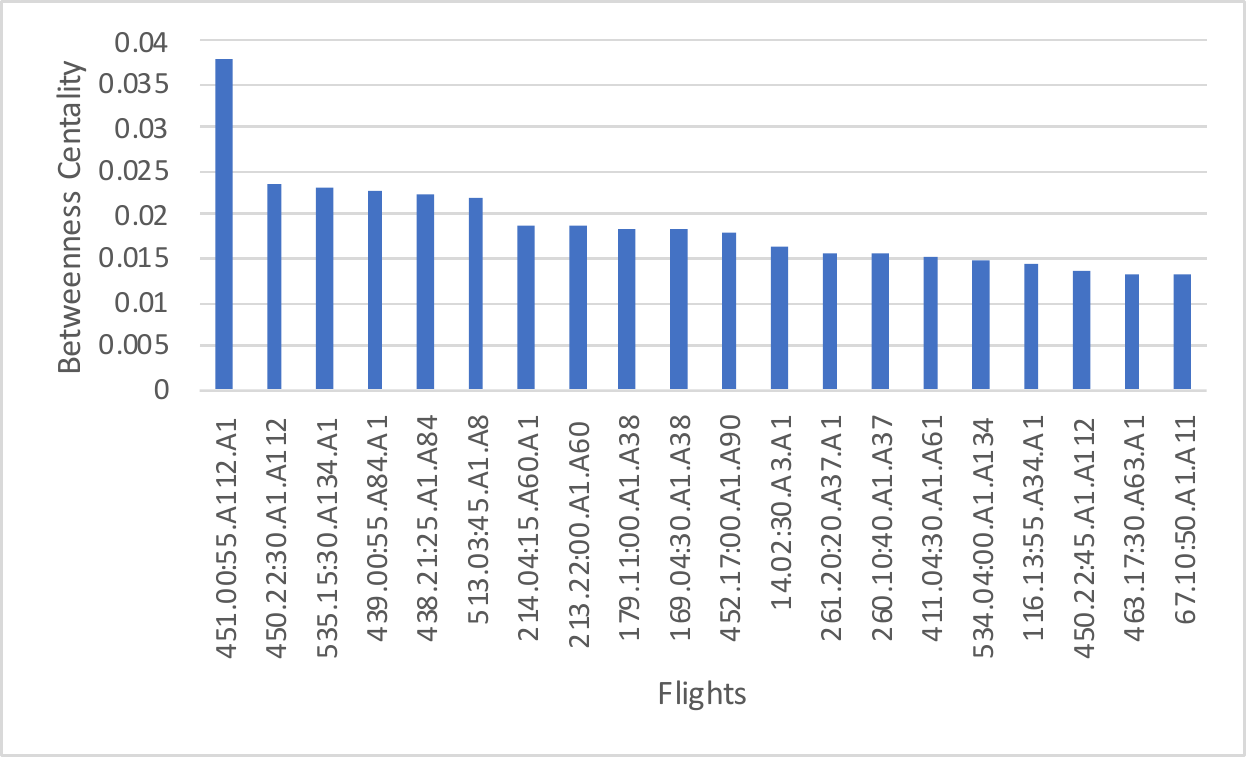}
			\caption{}
			\label{subfig_BF_B_FCDN}
		\end{subfigure}
		\caption{Disruptive flights using the node-betweenness centrality in FCN. (a) Connection network; (b) Delay network.}
		\label{subfig_BF_B_FC}
	\end{figure*}

\begin{figure*}[htb!]
		\centering
		\begin{subfigure}[t]{0.5\textwidth}
			\centering
			\includegraphics[width=\linewidth]{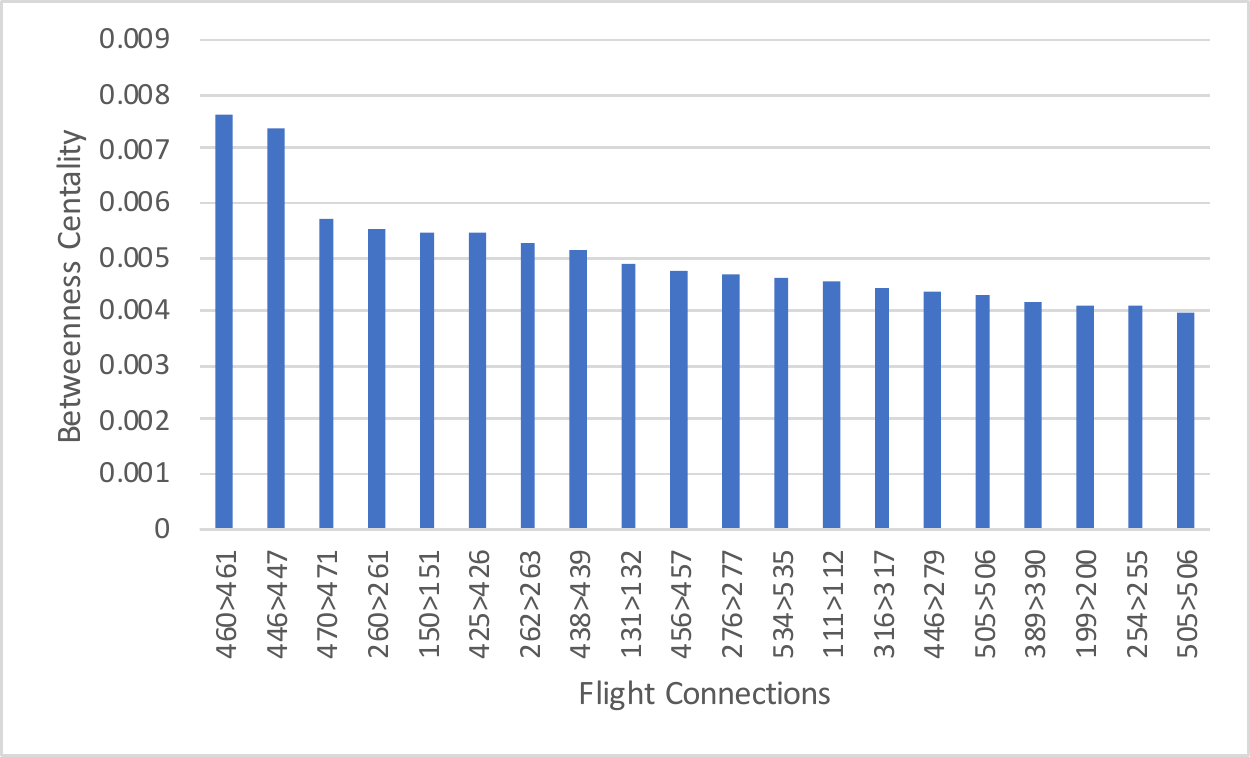}
			\caption{}
			\label{subfig_BC_B_FCN}
		\end{subfigure}%
		~ 
		\begin{subfigure}[t]{0.5\textwidth}
			\centering
			\includegraphics[width=\linewidth]{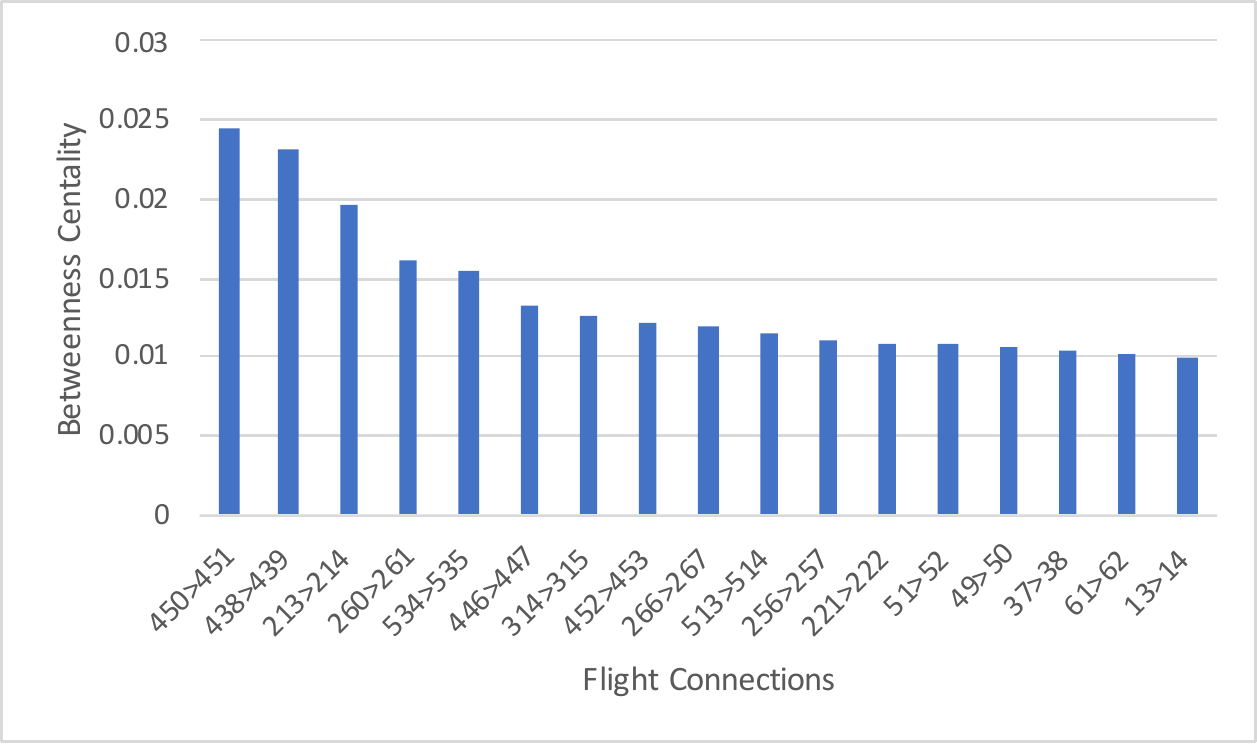}
			\caption{}
			\label{subfig_BC_B_FCDN}
		\end{subfigure}
		\caption{Disruptive connections using edge-betweenness centrality in FCN. (a) Connection network; (b) Delay network.}
		\label{fig_BC_B_FC}
	\end{figure*}

\begin{figure*}[htb!]
		\centering
		\begin{subfigure}[t]{0.5\textwidth}
			\centering
			\includegraphics[width=\linewidth]{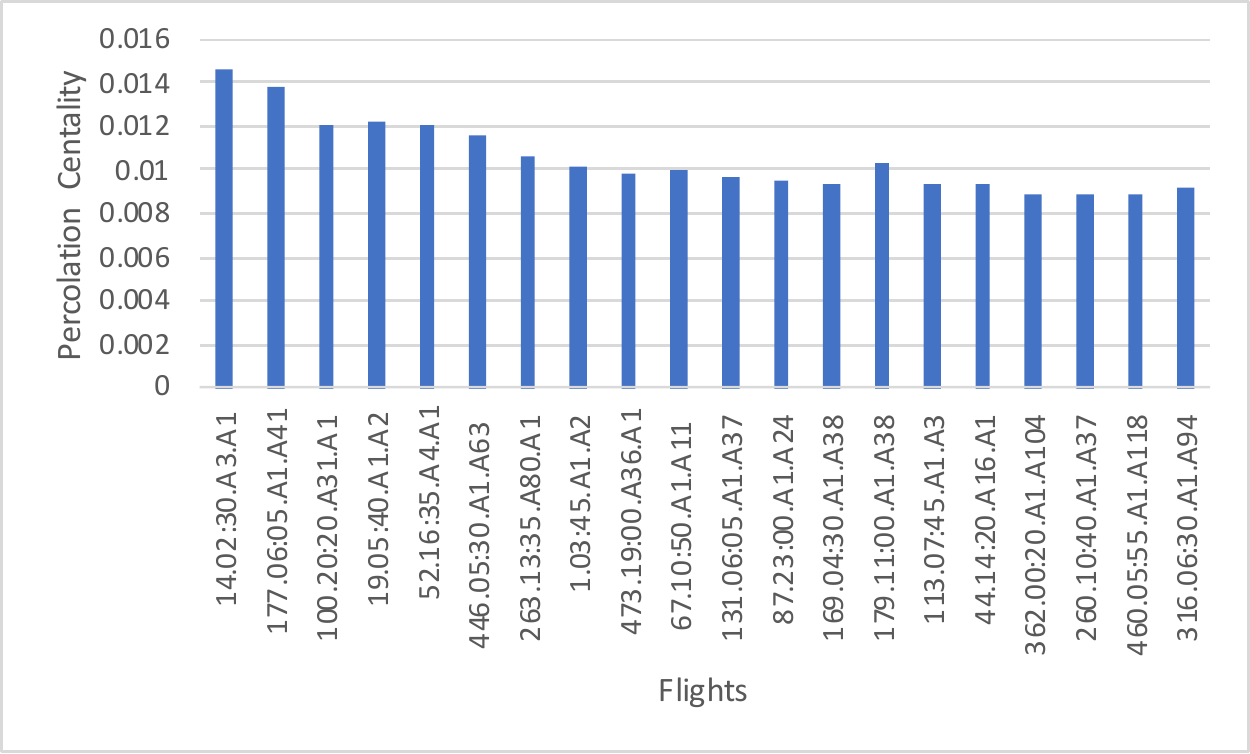}
			\caption{}
			\label{subfig_BF_P_FCN}
		\end{subfigure}%
		~ 
		\begin{subfigure}[t]{0.5\textwidth}
			\centering
			\includegraphics[width=\linewidth]{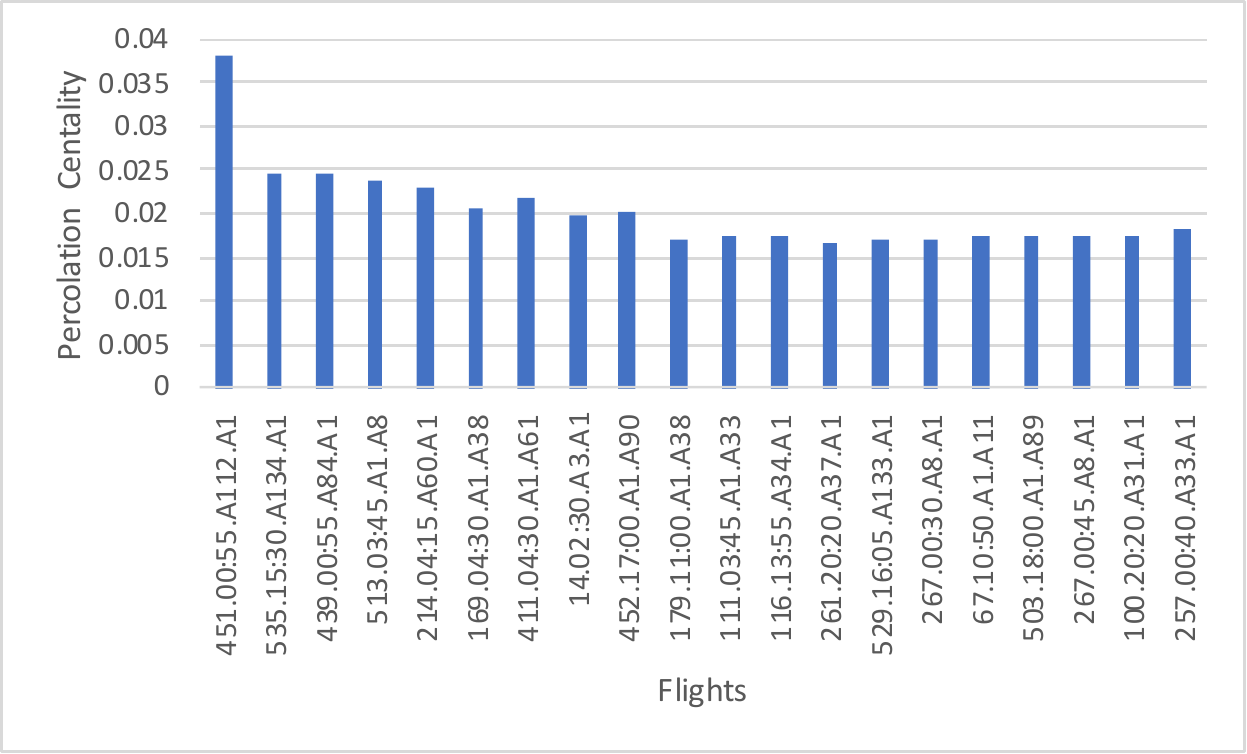}
			\caption{}
			\label{subfig_BF_P_FCDN}
		\end{subfigure}
		\caption{Disruptive flights using percolation in FCN. (a) Connection network; (b) Delay network.}
		\label{fig_BF_P_FC}
	\end{figure*}

	The results of the FCN are presented in Figs. ~\ref{subfig_FC}--\ref{fig_BF_P_FC}. Fig.~\ref{subfig_FC} shows the degree distribution, which is similar to the AFN, but it is quite different from real-life problems with a large number of peaks at the tail for both connection and delay networks. The degree distribution for the CN shows that most flights can affect only one or two flights, and the degree distribution for the delay network shows that most delays are propagated to one or two flights, that is the schedules of the airline are quite robust. We can observe some peaks in the tail of the degree distribution for the CN, which shows that some flights can influence many  flights because they have connections to several other flights; for the delay network, it shows that in some cases delays were actually propagated to many flights.
	
	Fig.~\ref{fig_BF_D_FC} represents the top disruptive flights using out-degree centrality, where Fig.~\ref{fig_BF_D_FC}(a) shows the flights that can influence many flights because they have many connections, according to the schedule. Fig.~\ref{fig_BF_D_FC}(b) shows the flights that actually lead to delayed propagation to many flights. There are some overlaps between Fig.~\ref{fig_BF_D_FC}(a) and (b), for example, flight numbers 451 and 457, which means that these flights were expected to cause delay propagation as per the schedule, and they actually led to delayed propagations.
	
	Fig.~\ref{subfig_BF_B_FC} represents disruptive flights as per the node-betweenness centrality, where Fig.~\ref{subfig_BF_B_FC}(a) shows flights with high connectivity through them. These flights are the potential disruptive elements because if anything goes wrong with them, they can influence several other flights. Fig.~\ref{subfig_BF_B_FC}(b) shows top flights through which delays were propagated to many flights. Flight numbers 179, 169, and 14 are common in Fig.~\ref{subfig_BF_B_FC}(a) and Fig.~\ref{subfig_BF_B_FC}(b), which means that as per the airline schedule, they were potential disruptive elements and actual disruptive elements when the schedule was executed; therefore, these flights need special attention to improve airline operation.
	
	Fig.~\ref{fig_BC_B_FC} represents the disruptive flight connections using the edge-betweenness centrality. Fig.~\ref{fig_BC_B_FC}(a) shows flight pairs, that is, connections through which there is the highest traffic flow and contain potential disruptive elements because delay propagation on these connections can cause delay propagation to several other flights. Fig.~\ref{fig_BC_B_FC}(b) shows connections that were delayed and led to the highest delay propagation to other flights. Connections (438,439), (446,447), and (260,261) are common to both potential and actual disruptive elements; therefore, attention is required to improve airline operations. To keep the figures clean, the labels on the axis were shortened by keeping only the flight number.
	
	Percolation is an interesting phenomenon in network science and it can be very useful in DNs to avoid large disruptions (Table~\ref{tab_properties}). We applied node percolation to the DN, and the results are represented in Fig.~\ref{fig_BF_P_FC}. This tells us which flights should avoid delays to avoid delay propagation and become a big disruption. This can be observed as a dynamic betweenness centrality, that is, removing the most influential flight using betweenness centrality scores, and then recalculating the centrality scores and removing the flight with the highest score and so on. This provides better results than betweenness centrality because when a flight is removed from the network, the connectivity of the remaining flights changes; thus, the second-highest flight in the first calculation of centrality may not be the same after calculating the new centralities. Notably, the order of flights represents the flights according to their importance, with the most influential flight on the left-hand side; however, their centralities are not in order. This is because the centralities are calculated for networks of different sizes because of the removal of flights. Flight numbers 14, 169, and 100 are potential and actual disruptive elements that require attention.
	
	From these results for FCN, it is evident that there are some overlaps in the disruptive elements of the CN and DN, which means that some flights could cause delays owing to their tight schedules, and these flights lead to delay propagation for other flights. Therefore, the airlines should pay attention to overlapped disruptive elements to improve their operations. In addition, some flights are common to different results; for example, flight 14 is an actual disruptive element from Figs. ~\ref{fig_BF_D_FC}–\ref{fig_BF_P_FC}, such flights require the most attention to improve operations. Moreover, not all potentially disruptive flights are actual disruptive elements, for example, flight number 390 in Fig.~\ref{fig_BF_D_FC}(a). This could be attributed to multiple reasons. For example, flights might have sufficient slack time between the connections; therefore, the delays from the incoming delayed flights are absorbed by slack time, or there could be alternate crew and aircraft arrangements to tackle the late connections. It is difficult to determine exactly what happened to  the flight at the time of operation from the available data because airlines use different recovery strategies. However, from the data, we observe that flight 390 had many potential connections, most of which were passenger connections, and it had sufficient slack times, of which only a few led to actual delay propagations; therefore, flight 390 is not an actual disruptive element.

	\subsection{Multilayer network}
	\label{subsec_MLN}
\begin{figure*}[htb]
		\centering
		\begin{subfigure}[t]{0.5\textwidth}
			\centering
			\includegraphics[width=\linewidth]{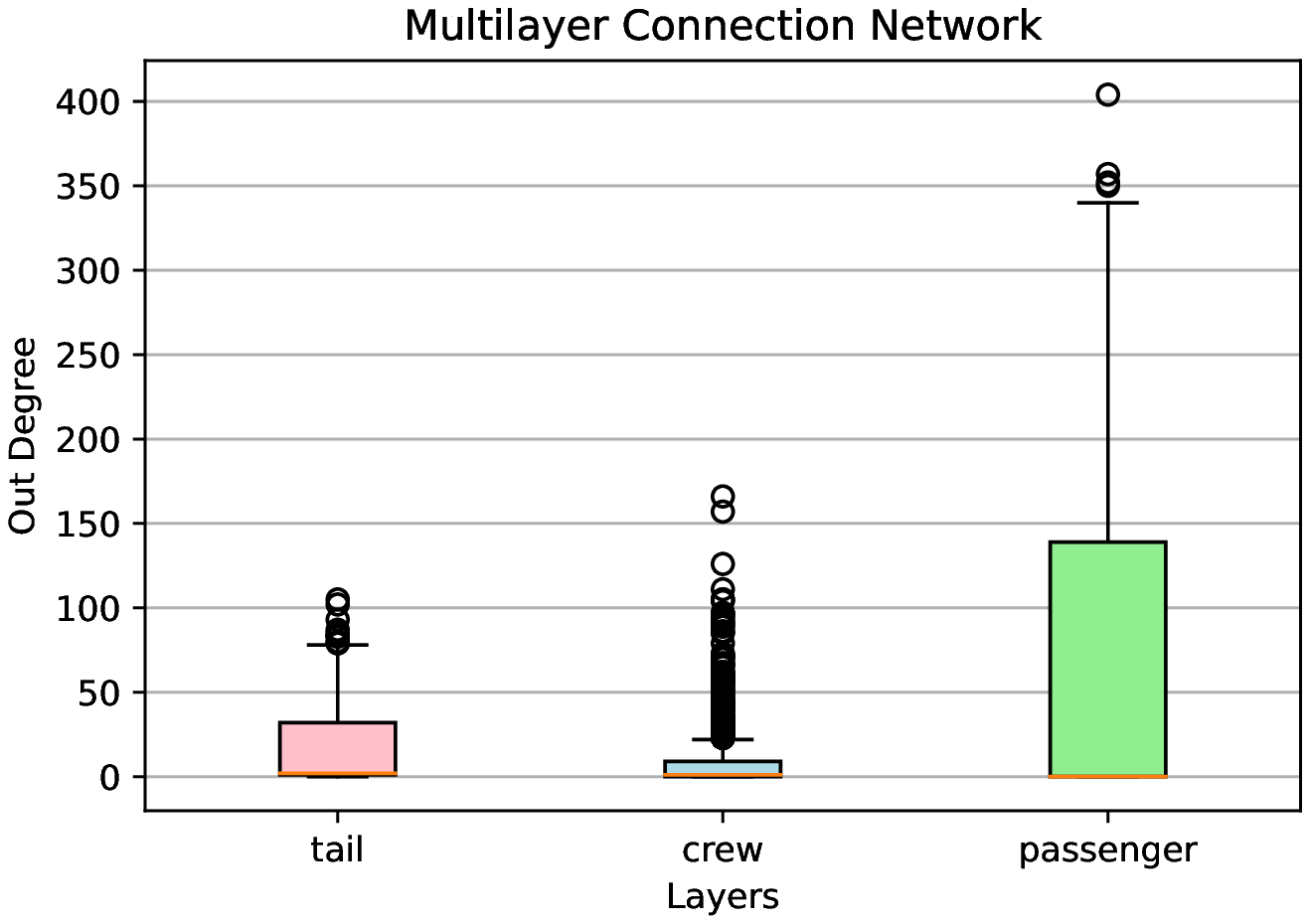}
			\caption{}
			\label{subfig_MLN_CN}
		\end{subfigure}%
		~ 
		\begin{subfigure}[t]{0.5\textwidth}
			\centering
			\includegraphics[width=\linewidth]{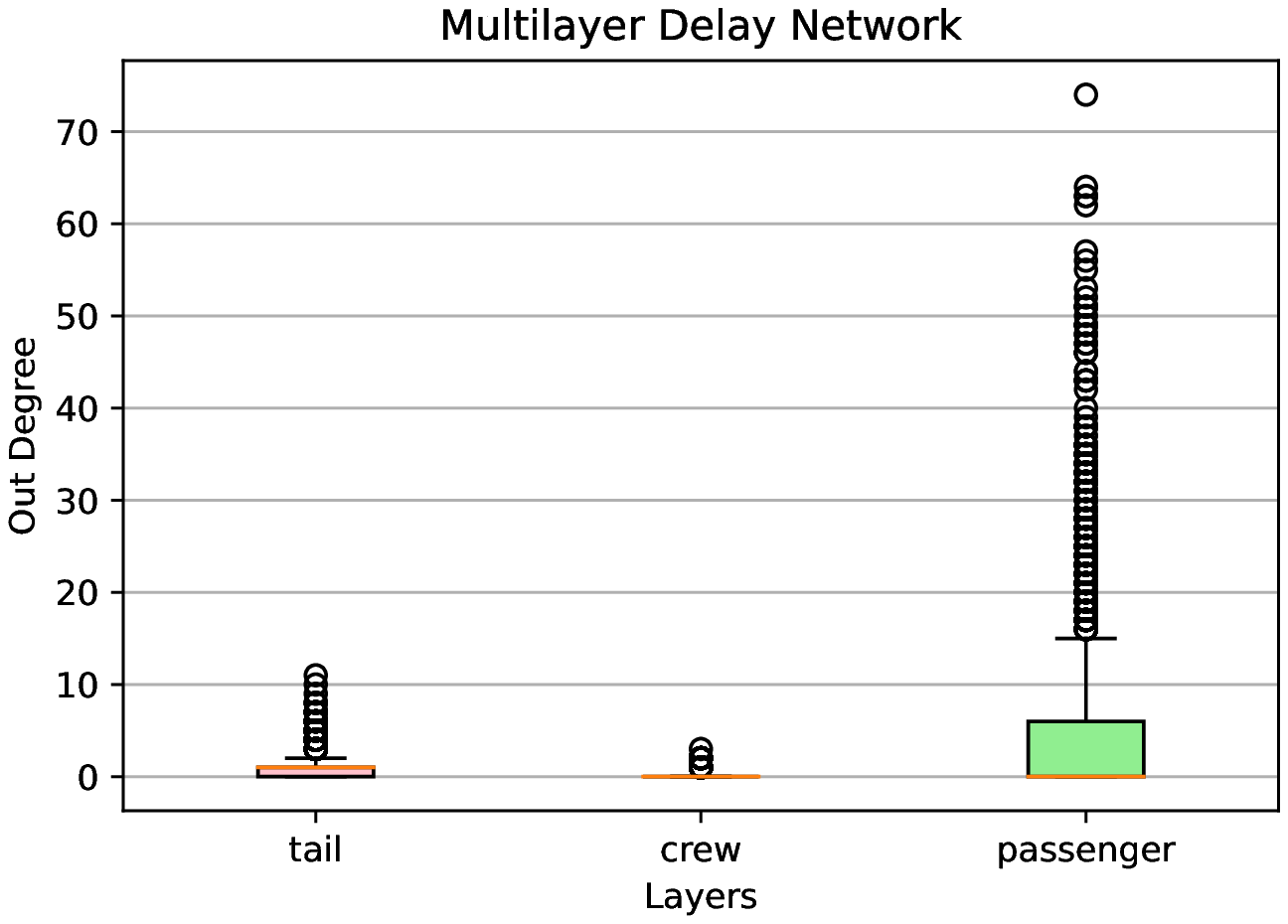}
			\caption{}
			\label{subfig_MLN_DN}
		\end{subfigure}
		\caption{MFCN out-degree distributions. (a) Connection network; (b) Delay network.}
		\label{fig_MLNet}
	\end{figure*}

	The results for the MFCN are represented in Fig. ~\ref{fig_MLNet} and Table~\ref{tab_mln_results}. Fig.~\ref{fig_MLNet} shows the degree distribution for different layers of multilayer connections and DNs using box plots. From Fig.~\ref{fig_MLNet}(a) and Fig.~\ref{fig_MLNet}(b), it is evident that there is greater variability in the mean of passenger connections as well as larger outliers for both the CN and DN. This means that passenger connections are not only potential causes for big disruptions but are also the actual cause for the biggest disruptions in the airline network. Tail connections are the second major disruptive element, and crew connections do not lead to large disruptions. Therefore, passenger connections are disruptive (both potential and actual) in airline networks.
	
	\begin{table}[htb]
		\centering
		\caption{Comparative study of different connection types of MFCN.}
		\label{tab_mln_results}
		{\small \begin{tabular}{llllll}
				\hline
				\multicolumn{2}{c}{\textbf{Property}}                                           & \multicolumn{1}{c}{\textbf{all}} & \multicolumn{1}{c}{\textbf{tail}} & \multicolumn{1}{c}{\textbf{crew}} & \multicolumn{1}{c}{\textbf{pax}} \\ \hline
				\multirow{2}{*}{Density}                                                      & CN & 0.063292               & 0.013347               & 0.007182               & 0.053296              \\
				& DN & 0.005596               & 0.000986                & 0.000182                & 0.004820                \\ \hline
				\multicolumn{6}{l}{%
					\begin{minipage}{10cm}%
						\footnotesize Note: CN: Connection Network, DN: Delay Network
					\end{minipage}%
				}\\	
		\end{tabular}}
	\end{table}
	
	Table~\ref{tab_mln_results} compares the network properties for the different connection types of the network. The table shows that the density of passenger connections in the CN is greater than that of crew and tail connections, which means that the airline has more passenger connections than crew and tail connections. Similarly, the density of the passenger connection is greater than that of the  crew and tail in the DN, which means that most of the delays are propagated through passenger connections. Therefore, passenger connections are the most disruptive connection type for the airline, as already pointed out by the degree distribution; thus, airlines need to pay attention to passenger connections to avoid delay propagations .

	\section{Conclusions and implications}
	\label{sec_conclusion}
	Here, we discuss the theoretical and managerial implications, limitations, and future scope of study.
	
	\subsection{Implications for research}
	We propose modeling airline schedules and historical operational data as AFN, FCN, and MFCN, each with two variants: CN and DN. We applied network science concepts and techniques such as centrality, percolation, and MNs to find disruptive airports, flights, flight -connections and connection-types. Connection networks model airline schedules as networks and provide potentially disruptive elements, whereas delay networks model the historical operational data and provide actual disruptive elements. The proposed work analyzes the historical operational data and the airline schedules, and cross-checks for disruptive elements in the schedule and historical operational data. A ranking system is also proposed to identify and prioritize the potential and actual disruptive elements obtained from different modeling and network science techniques. The unique modeling approach and ability to directly identify disruptive elements, such as airports, flights, flight connections, and connection types, distinguishes the proposed methodology from existing work, for example, \cite{Ledwoch2022,Gui2020,Wong2012,Wu2019}.
	
	The proposed idea is validated with a case study of an airline, and it is observed that there is some overlap between potential and actual disruptive elements, which indicates that some issues with airline schedules itself cause delay propagations. The airline network has a small-world effect with a diameter of four, which means that the airline has high connectivity, which can result in delays propagating from any part of the network to any other part with only four flight delays. As reported in \cite{Ledwoch2022}, passenger flight connections from one flight to another cause the highest number of delays.
	
	There are two types of disruptive elements: potential disruptive elements analyzed from airline schedules and actual disruptive elements analyzed from historical operational data, which should be interpreted and prioritized appropriately. The disruptive elements in schedules and historical data indicate that an issue in the airline schedule could cause flight delays and has the highest priority. Similarly, disruptive elements found in DNs using different techniques, such as percolation and betweenness centrality, have a high priority. However, some disruptive elements are least important, such as those occurring only in schedules, and they could be extensively used.
	
	Thus, the theoretical or methodological contribution of this study to solve the delay propagation problem is the development of a novel network science-based methodology to model airline schedules and historical operational data that can identify the most disruptive airports, flight numbers, flight connections, and flight connection types. This study also discusses guidelines for interpreting and prioritizing disruptive elements. Thus, the proposed methodology provides airlines with a simple tool for building robust flight schedules by focusing on the disruptive elements. This could help reduce the negative impacts of delay propagation on economic losses, reputation of airlines, time, and money of passengers, and could have a positive environmental impact. Finally, the analysis can be extended to other network analyses to identify key elements, such as rail  and telecommunication networks.
	
	\subsection{Implications for management practice}
	From a managerial perspective, the study helps an airline in analyzing their operations, build robust flight schedules, and best utilize its resources. First, it helps evaluate schedules and improve them by looking for overlapping disruptive elements in CN and DNs, which indicates issues with the schedule. Second, it helps an airline look at historical operational data and identify the culprits of big disruptions by looking for disruptive elements and paying attention to them in the future to avoid the same disruptions again. Airlines can deal with disruptive elements in different ways; for example, they can introduce extra slack time for disruptive passenger connections, arrange extra crew for disruptive crew connections, and arrange extra aircraft for disruptive tail connections. Hub airports also require special attention as they are the busiest and one of the most disruptive elements. Similarly, for disruptive flights, there should be sufficient slack time or extra crew or tail arrangements, depending on the delay codes of the disruptive element. Thus, the analysis provides a tool for improving the operations of an airline and it can be applied as follows. First, the analysis is run to identify potential and actual disruptive elements using different network science techniques, rank the disruptive elements (as described in Section~\ref{sec_analysis}) according to their priority, and attend to them to improve the schedule and operations of an airline.
	
	The data required for using the proposed methodology are captured by all airlines in their software systems; thus, the proposed methodology and the tool for carrying out the analysis can be integrated with the main systems of the airlines. This can perform automated analysis for the airlines and can identify top (`N’ number of) disruptive elements as per their ranking. Thus, an automated and integrated system could be developed to manage and improve airline operations.
	
	Although the full potential of the proposed network science-based approach is achieved by having complete data, publicly available airline data can be used to obtain some of the results. Another managerial use case for an airline is the development of a benchmarking system to compare the performance of the airline against competitors using publicly available data from public organizations, such as the Bureau of Transportation Statistics in the U.S. States or aggregators (such as FlightStats and FlightRadar24). Moreover, publicly available airline data can be augmented with data simulators, such as \cite{barnhart2014modeling}, to perform the analysis.

	\subsection{Limitations and future research}
	In this study, we proposed a novel network science-based methodology to solve the delay propagation problem for an airline’s flight network by identifying the top disruptive elements and demonstrated it using a case study of an airline. This methodology provides airlines with an important tool for improving operations by identifying the most disruptive elements of airline networks. However, the proposed network science methodology can be extended to other fields, where data can be presented in terms of networks and have events such as disruptions. For example, the analysis can be adapted to other transportation networks, such as rail networks, which can help identify the most disruptive trains or routes. Similarly, the analysis can be extended to telecommunication networks to identify key elements of the network, for example, the busiest routers and channels; accordingly, extra resources can be added or alternatives can be kept to handle disruptive situations.
	
	Disruptive element analysis presents a static view of airline operations over time, which is helpful for the airline to reflect upon, analyze its performance, and make decisions to improve operations. However, the airline schedule can be further improved and made more robust to disruptions by closely analyzing the operations of the airline on an hourly basis. Thus, possible extension of the work is dynamic analysis of the disruptive elements in the airline networks, to improve the operations further and will be studied in the future extension of the work.
	
	\section*{Acknowledgement}
	This work is part of a BOEING project `Airline Performance and Disruption Management Across Extended Networks (APEMEN)’ funded with research grant number 46599. We also want to thank the anonymous reviewers for their valuable feedback, which has greatly improved the paper.
	
	\subsection*{Declaration of competing interest}
	The authors declare that there are no conflicts of interest.	
	\printbibliography
	
\end{document}